\documentclass[apjl]{emulateapj}

\usepackage{graphics}
\usepackage{mathptmx}
\usepackage{natbib}

\newcommand{\teff}{$T_{\!\mbox{\tiny\em eff}}$}
\newcommand{\logg}{$\log\,g$}

\newcommand{\oh}{12\,+\,log(O/H)~=\/}

\newcommand{\hi}{H\,{\sc i}\rm}
\newcommand{\hii}{H\,{\sc ii}\rm}
\newcommand{\hei}{He\,{\sc i}\rm}
\newcommand{\heii}{He\,{\sc ii}\rm}
\newcommand{\oiii}{[O\,{\sc iii}]}
\newcommand{\oii}{[O\,{\sc ii}]}
\newcommand{\sii}{[S\,{\sc ii}]}

\newcommand{\neiii}{[Ne\,{\sc iii}]}

\newcommand{\caii}{Ca\,{\sc ii}}
\newcommand{\cai}{Ca\,{\sc i}}
\newcommand{\siii}{Si\,{\sc ii}}
\newcommand{\siiii}{Si\,{\sc iii}}
\newcommand{\siiv}{Si\,{\sc iv}}

\newcommand{\nii}{N\,{\sc ii}}
\newcommand{\niii}{N\,{\sc iii}}
\newcommand{\cii}{C\,{\sc ii}}
\newcommand{\ciii}{C\,{\sc iii}}
\newcommand{\civ}{C\,{\sc iv}}
\newcommand{\mgii}{Mg\,{\sc ii}}
\newcommand{\fei}{Fe\,{\sc i}}
\newcommand{\feii}{Fe\,{\sc ii}}
\newcommand{\tiii}{Ti\,{\sc ii}}
\newcommand{\crii}{Cr\,{\sc ii}}
\newcommand{\ovi}{O\,{\sc vi}}
\newcommand{\ov}{O\,{\sc v}}
\newcommand{\ariv}{[Ar\,{\sc iv}]}

\newcommand{\te}{$T_e$}
\newcommand{\toiii}{$T$(O\,{\sc iii})}
\newcommand{\toii}{$T$(O\,{\sc ii})}

\newcommand{\lin}{$\,\lambda$}
\newcommand{\llin}{$\,\lambda\lambda$}

\slugcomment{}

\shorttitle{Blue supergiants in IC 1613}
\shortauthors{Bresolin et al.}

\begin{document}


\title{VLT spectroscopy of blue supergiants in IC~1613$^1$\\[2mm]}\footnotetext[1]{Based on VLT observations for ESO Large Programme 171.D-0004.}

\author{Fabio Bresolin and Miguel A. Urbaneja} \affil{Institute for Astronomy, 2680 Woodlawn Drive, Honolulu, HI 96822;\\ bresolin@ifa.hawaii.edu, urbaneja@ifa.hawaii.edu}

\and

\author{Wolfgang Gieren and Grzegorz Pietrzy\'nski} \affil{Universidad de Concepci\'on, Departamento de F\'isica, Casilla 160-C, Concepci\'on, Chile; \\ wgieren@astro-udec.cl, pietrzyn@hubble.cfm.udec.cl}

\and
 
\author{Rolf-Peter Kudritzki} \affil{Institute for Astronomy, 2680 Woodlawn Drive, Honolulu, HI 96822;\\ kud@ifa.hawaii.edu}

\begin{abstract}

We present multi-object spectroscopy of young, massive stars in the Local Group galaxy IC~1613. We provide the spectral classification and a detailed spectral catalog for 54 OBA stars in this galaxy. The majority of the photometrically selected sample
is composed of B- and A-type supergiants. The remaining stars include early O-type dwarfs and the only Wolf-Rayet star known in this galaxy.
Among the early B stars we have serendipitously uncovered 6 Be stars, the largest spectroscopically confirmed sample of this class of objects beyond the Magellanic Clouds. We measure chemical abundances for 9 early-B supergiants, and find a mean oxygen abundance of 12~+~log(O/H)~=~7.90~$\pm$~0.08. This value is 
consistent with the result we obtain for two \hii\/ regions in which we detect the temperature-sensitive \oiii\lin4363 auroral line.

\end{abstract}

\keywords{galaxies: abundances --- galaxies: stellar content --- galaxies:
  individual (IC 1613) --- stars: early-type}
 

\section{Introduction}

\begin{figure*} \epsscale{1.} \plotone{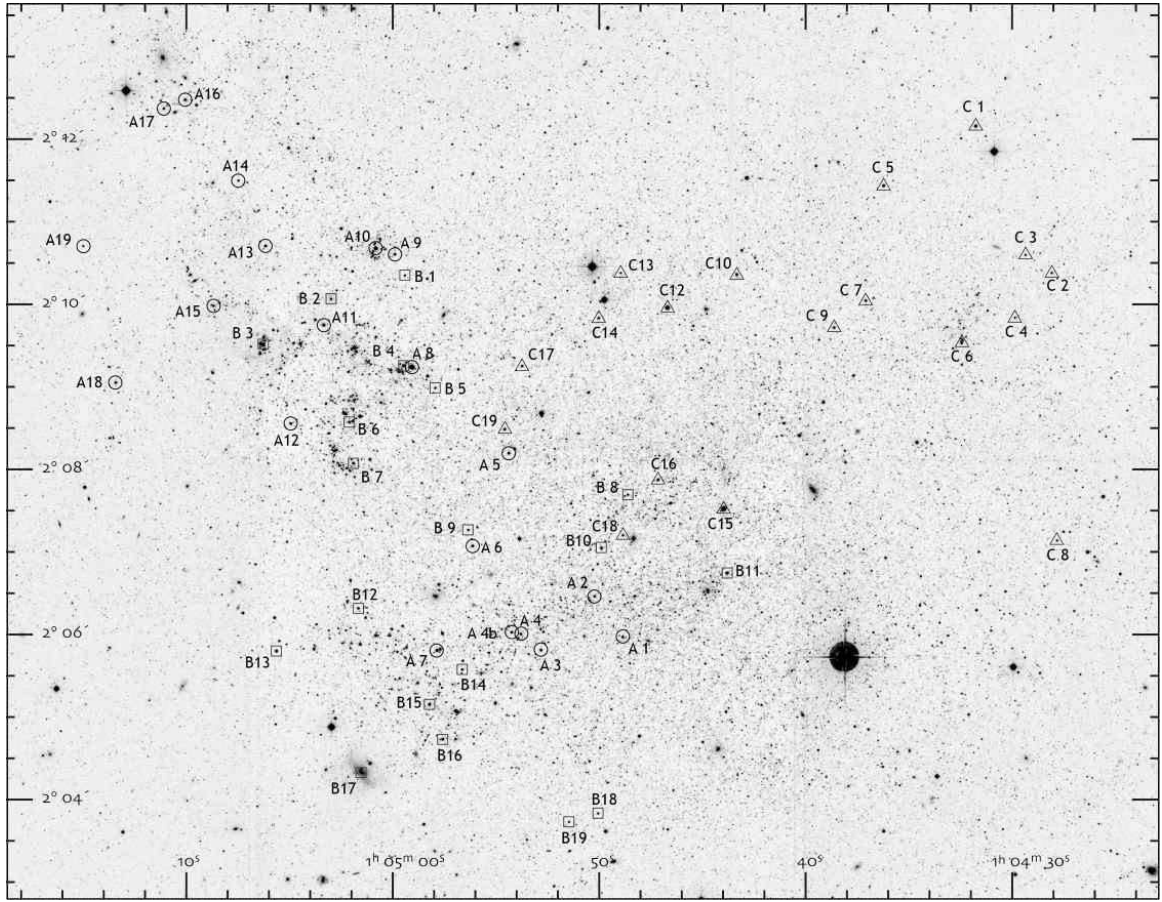} \caption{Finding chart for the
spectroscopic targets. Different symbols are used for stars in fields A
(circles), B (squares) and C (triangles). The image has been obtained with a
Sloan $g$ filter at the MegaPrime imager at the CFHT telescope on Mauna Kea.
North is up and east is to the left. (courtesy L.~Rizzi and B.~Tully)}
\label{chart} \end{figure*}

IC~1613 is a Local Group dwarf irregular galaxy, member of the M31 subgroup
(\citealt{McConnachie:2006}), located at a distance of $721\pm12$~kpc
(\citealt{Pietrzynski:2006}). Its low metal content has been derived both from
the photometry of the intermediate-age population of stars
([Fe/H]~$\simeq$~$-1.3$: \citealt{Cole:1999}, \citealt{Rizzi:2007}) and from
\hii\/ region spectra [12~+~log(O/H)~$\simeq$~7.70: \citealt{Kingsburgh:1995},
\citealt{Lee:2003}].

The evolved stellar populations of IC~1613 have been examined in detail from
Hubble Space Telescope imaging programs. The star formation rate in an outlying
field 1.6~kpc in projection from the galactic center was shown by
\citet{Skillman:2003} to have peaked between 3 and 6 Gyr ago, and during the
last Gyr it has been supressed if compared to the central field studied by
\citet{Cole:1999}. The ongoing star formation activity is concentrated in an
off-center region, $\sim$1~kpc to the north-east of the geometrical center of
IC~1613, in correspondence of the peak of the \hi\/ distribution
(\citealt{Lake:1989}) and of the largest concentration of \hii\/ region
complexes (\citealt{Hodge:1990}), giant ionized shells (\citealt{Meaburn:1988}),
young stars (\citealt{Hodge:1991}) and OB associations
(\citealt{Hodge:1978,Borissova:2004}).

The young stellar content of IC~1613 has been investigated by
\citet{Sandage:1976}, \citet{Freedman:1988} and \citet{Georgiev:1999}, among
others. More recently, \citet{Valdez-Gutierrez:2001} and \citet{Lozinskaya:2003}
 have examined the interplay between young stars and the kinematics of the
neutral and ionized gas, and the effects of stellar winds and supernova
explosions on the structure and energy balance of the ISM. The presence of
numerous OB associations appears to provide sufficient mechanical energy to
explain the observed expansion of most of the superbubbles within the standard
framework of \citet{Weaver:1977}. However, \citet{Silich:2006} find that the
multiple winds and supernovae model is not consistent with the observed mass and
dimensions of the largest, kpc-size \hi\/ supershell. The incomplete
characterization of the massive stellar population, in particular with regard to
the spectral types of the OB stars that ionize the gas and that blow their
stellar winds into the interstellar medium, are major limitations for these
studies on the links between young stars and gas in this galaxy.

Previous spectroscopic work on the stellar content of IC~1613 has focused on the
search for Wolf-Rayet (W-R) stars and other emission-line objects
(\citealt{Lequeux:1987}, \citealt{Azzopardi:1988}, \citealt{Armandroff:1991}).
The high-excitation WO3 star in the \hii\/ region S3 (\citealt{Sandage:1971}),
in particular, has received a great deal of attention since its discovery
(\citealt{Dodorico:1982}, \citealt{Davidson:1982}, \citealt{Garnett:1991},
\citealt{Kingsburgh:1995}), and remains the only confirmed W-R star in this
galaxy. In addition to the early work by \citet{Humphreys:1980}, who obtained
spectra of 7 of the brightest BA and M supergiants with the Kitt Peak 4m
telescope, the spectroscopy of a dozen stars in the NE quadrant of IC~1613
obtained with the Special Astrophysical Observatory 6m telescope has been
published by \citet{Lozinskaya:2002}. However, the quality of these spectra is
only adequate for a preliminary spectral classification.

In this paper we present 3-5~\AA\/ resolution spectra of 54 OBA stars in IC~1613
brighter than $V=20.2$, obtained with the European Southern Observatory (ESO)
Very Large Telescope (VLT) in the context of the Araucaria Project
(\citealt{Gieren:2005}). This work follows similar spectroscopic work recently
carried out by our group on blue supergiants in the galaxies NGC~300
(\citealt{Bresolin:2002}), WLM (\citealt{Bresolin:2006}) and NGC~3109
(\citealt{Evans:2007}). One main motivation for this paper is to present a
catalog of spectra and a classification of spectral types for the observed
targets. This will be useful for future spectroscopic investigations of the blue
supergiants in IC~1613 at higher spectral resolution. Moreover, we carry out the
first determination of the chemical abundances of blue supergiants in this
galaxy, by fitting model spectra to the data obtained for 9 early B-type stars.
The paper is organized as follows: in \S2 we describe the observations and the
data reduction, and in \S3 we present the spectral catalog. In \S4 and \S5 we
discuss the chemical abundances measured in B-type stars and \hii\/ regions,
respectively. We conclude with a summary in \S6.

\section{Observations}

\begin{deluxetable*}{rrrrrcccccl}[h] 
\tabletypesize{\scriptsize}
\tablecolumns{11}
\tablewidth{0pt}
\tablecaption{IC 1613 - Spectroscopic targets\label{catalog}}

\tablehead{
\colhead{}     &
\colhead{R. A.}            &
\colhead{Decl.}           &
\colhead{}             &
\colhead{}     &
\colhead{}    &
\colhead{}			& 
\colhead{$W_\gamma$} &
\colhead{} &
\colhead{$V_{\rm Helio}$} &
\colhead{}\\
\colhead{Slit Number}   &
\colhead{(J2000.0)}         &
\colhead{(J2000.0)}         &
\colhead{$V$}         &
\colhead{$V-I$}         &
\colhead{Spectral Type}		&
\colhead{K/(H + H$\epsilon$) }			&
\colhead{(\AA)}			&
\colhead{S/N}	&
\colhead{(km\,s$^{-1}$)}	&
\colhead{Comments\tablenotemark{a}}\\[1mm]
\colhead{(1)}	&
\colhead{(2)}	&
\colhead{(3)}	&
\colhead{(4)}	&
\colhead{(5)}	&
\colhead{(6)}	&
\colhead{(7)}	&
\colhead{(8)}	&
\colhead{(9)}	&
\colhead{(10)}	&
\colhead{(11)}}
\startdata
\\[-6.13mm]
\cutinhead{Field A}
1\dotfill   & 1 04 48.89  &   2 05 59.5 &     18.81 & 0.17 & 	A0~II      & 0.27     & 5.4    	&  74 & $-229$ & \hfill SK~E33\\
2\dotfill   & 1 04 50.26  &   2 06 29.2 &     19.48 & 0.38 & 	A5~II      & 0.81     & 6.8     &  54 & $-214$ & \hfill SK~E42\\
3\dotfill   & 1 04 52.87  &   2 05 50.1 &     18.55 & 0.44 & 	A7-F0~Ib   & 0.95     & 5.4     &  77 & $-235$ & \hfill SK~F33\\
4\dotfill   & 1 04 53.84  &   2 06 01.9 &     19.05 & 0.08 & 	B9~II      & 0.14     & 4.7     &  68 & $-225$ & \hfill SK~23a\\
4b\dotfill  & 1 04 54.29  &   2 06 03.1 &     18.98 & $-$0.08 &   B5~Ib      & 0.20     & 3.1     &  52 & $-276$ & \hfill SK~23A\\
5\dotfill   & 1 04 54.41  &   2 08 13.7 &     18.61 & $-$0.11 & 	B3~Ib  	   & 0.13     & 2.6     &  87 & $-230$ &  \\
6\dotfill   & 1 04 56.17  &   2 07 06.1 &     18.66 &  0.19 & 	A3~II      & 0.59     & 6.4     &  81 & $-228$ & \\
7\dotfill   & 1 04 57.92  &   2 05 49.9 &     18.41 & $-$0.11 & 	B2~Iab 	   & 0.13     & 2.2     &  96 & $-250$ &  \\
8\dotfill   & 1 04 59.06  &   2 09 17.0 &     16.38 & 0.10 & 	A2~Ia      & 0.44     & 0.41    & 235 & $-241$ & {\scriptsize \sc hii} \hfill SK~A43 \\
9\dotfill   & 1 04 59.94  &   2 10 38.6 &     19.19 & $-$0.16 & 	B5~Iab 	   & 0.17     & 2.6     &  72 & $-223$ & {\scriptsize \sc hii} \hfill SK~22c \\
10\dotfill  & 1 05 00.87  &   2 10 43.0 &     17.46 & $-$0.15 & 	B1~Ia  	   & 0.20     & 1.4     & 159 & $-255$  & {\scriptsize \sc hii} \hfill SK~22A\\
11\dotfill  & 1 05 03.39  &   2 09 47.2 &     18.08 & 0.07 & 	B9~Ia 	   & 0.30     & 1.7     & 110 & $-241$  & {\scriptsize \sc hii}\\
12\dotfill  & 1 05 04.99  &   2 08 35.4 &     18.78 & $-$0.14 & 	B1.5~Iab   & 0.17      & 2.1     &  89 & $-226$ & {\scriptsize \sc hii} \hfill SK~C52\\
13\dotfill  & 1 05 06.21  &   2 10 44.8 &     19.02 & $-$0.24 & 	O3-O4~V((f)) & 0.23     & 2.2     &  82 & $-240$  & {\scriptsize \sc hii} \hfill SK~22b \\
14\dotfill  & 1 05 07.51  &   2 11 32.4 &     19.85 & 0.13 & 	Be       & $<0.25$   & in em.  & 52 & $-210$  & \\
15\dotfill  & 1 05 08.74  &   2 10 01.1 &     19.35 & $-$0.24 & 	O9.5~III    & 0.32      & 1.7     &  74 & $-279$ & {\scriptsize \sc hii}\\
16\dotfill  & 1 05 10.10  &   2 12 31.3 &     18.90 & $-$0.18 & 	early B~Ib & \nodata   & 2.9     &  86 & $-233$  & \\
17\dotfill  & 1 05 11.13  &   2 12 25.0 &     19.33 & 0.21 & 	Be       & \nodata   & in em.  &  65 & $-195$  & \\
18\dotfill  & 1 05 13.50  &   2 09 05.3 &     19.23 & $-$0.18 & 	B1~Iab     & 0.17      & 2.3      & 70 & $-229$  & {\scriptsize \sc hii}\\
19\dotfill  & 1 05 15.05  &   2 10 45.0 &     20.19 & 0.14 & 	Be       & 0.73      & in em.   & 43 & $-206$  & \\
\cutinhead{Field B}
1\dotfill   & 1 04 59.44  &	2 10 23.2 &     19.57 & 0.04 & 	B9~II      & 0.23     & 6.3    	&  60 & $-253$ & {\scriptsize \sc hii} \hfill SK~A19\\
2\dotfill   & 1 05 03.03  &	2 10 06.1 &     19.68 & $-$0.27 & 	O5-O6~V    & 0.34     & 1.2     &  66 & $-229$ & {\scriptsize \sc hii} \hfill S17\tablenotemark{b}\\
3\dotfill   & 1 05 06.34  &	2 09 32.9 &     17.81 & $-$0.14 & 	B0~Ia      & 0.12     & 1.4     & 154 & $-250$ & {\scriptsize \sc hii} \hfill SK~B42\\
4\dotfill   & 1 04 59.52  &	2 09 17.6 &     18.27 & $-$0.20 & 	B1.5~Ia    & 0.19     & 1.8     & 114 & $-243$ & {\scriptsize \sc hii} \hfill SK~A42\\
5\dotfill   & 1 04 57.98  &	2 09 01.2 &     19.61 & 0.03 & 	Be  	   & \nodata  & in em.  &  57 & $-257$ & {\scriptsize \sc hii} \hfill SK~32b \\
6\dotfill   & 1 05 02.16  &	2 08 36.5 &     18.82 & $-$0.05 & 	B1~Ia      & 0.19     & 1.5     &  92 & $-268$ & {\scriptsize \sc hii} \hfill SK~38a\\
7\dotfill   & 1 05 01.95  &	2 08 06.5 &     18.99 & $-$0.29 & 	O9~I 	   & 0.14     & 2.2     &  87 & $-241$ & {\scriptsize \sc hii} \hfill SK~38b \\
8\dotfill   & 1 04 48.63  &	2 07 43.3 &     19.50 & $-$0.01 & 	B8~Ib      & 0.10     & 3.6     &  63 & $-249$ & \hfill SK~D31 \\
9\dotfill   & 1 04 56.38  &	2 07 17.8 &     19.72 & 0.18 & 	A3~II 	   & 0.57     & 7.7     &  53 & $-232$ &  \\
10\dotfill  & 1 04 49.93  &	2 07 04.7 &     18.94 & $-$0.01 & 	B8~Iab     & 0.10     & 3.3     &  84 & $-236$ & \hfill SK~E46\\
11\dotfill  & 1 04 43.82  &	2 06 46.1 &     18.68 & $-$0.16 & 	O9.5~I   & 0.23     & 2.2     &  93 & $-257$ & {\scriptsize \sc hii} \hfill SK~L\\
12\dotfill  & 1 05 01.72  &	2 06 20.7 &     19.03 & 0.13 & 	Be       & \nodata  & in em.  &  80 & $-225$ & \hfill SK~J46\\
13\dotfill  & 1 05 05.70  &	2 05 49.6 &     18.56 & $-$0.10 & 	B5~Iab     & 0.15     & 2.2     &  98 & $-221$ & \hfill SK~PeM \\
14\dotfill  & 1 04 56.69  &	2 05 36.0 &     19.32 & $-$0.23 & 	B0~Iab     & 0.08     & 2.2     &  74 & $-228$ & \hfill SK~H33\\
15\dotfill  & 1 04 58.28  &	2 05 10.5 &     18.52 & 0.14 & 	A0~II      & 0.31     & 5.6     &  98 & $-267$ & {\scriptsize \sc hii} \hfill SK~H34\\
16\dotfill  & 1 04 57.64  &	2 04 45.2 &     18.80 & $-$0.14 & 	B1.5~Iab   & 0.18     & 2.3     &  91 & $-242$ & {\scriptsize \sc hii} \hfill SK~H8\\
17\dotfill  & 1 05 01.62  &	2 04 21.1 &     19.90 & $-$0.13 & 	WO3         & \nodata  & in em.  &  47 & $-245$ & {\scriptsize \sc hii} \hfill DR1\tablenotemark{c} S3\tablenotemark{b}\\
18\dotfill  & 1 04 50.11  &	2 03 51.0 &     19.78 & 0.09 & 	B9~II      & 0.11     & 5.5      & 51 & $-250$ & \\
19\dotfill  & 1 04 51.52  &	2 03 44.8 &     19.94 & 0.01 & 	Be       & 0.37     & in em.   & 47 & $-245$ & \\
\cutinhead{Field C}
1\dotfill   & 1 04 31.75  &	2 12 11.1 &     18.11 & 0.11 & 	A2~II      & 0.34     & 6.1    	&  72 & $-269$ & \\
2\dotfill   & 1 04 28.07  &	2 10 24.0 &     19.58 & 0.17 & 	A3~II      & 0.61     & 7.0     &  35 & $-254$ & \\
3\dotfill   & 1 04 29.35  &	2 10 37.0 &     19.76 & 0.26 & 	A3~II      & 0.71     & 7.5     &  31 & $-259$ & \\
4\dotfill   & 1 04 29.87  &	2 09 51.5 &     20.14 & $-$0.08 & 	B8~II      & 0.15     & 4.7     &  28 & $-249$ & \\
5\dotfill   & 1 04 36.22  &	2 11 27.7 &     18.18 & 0.52 & 	A~V  	   & 0.81     & 3.2     &  57 & $17$ & \\
6\dotfill   & 1 04 32.45  &	2 09 33.1 &     18.83 & $-$0.07 & 	B1~Ia      & 0.22     & 1.5     &  57 & $-245$ & {\scriptsize \sc hii}\\
7\dotfill   & 1 04 37.09  &	2 10 04.0 &     19.04 & $-$0.24 & 	B1~Ia 	   & 0.20     & 1.9     &  48 & $-262$ &  \\
8\dotfill   & 1 04 27.84  &	2 07 09.2 &     19.90 & 0.18 & 	A3~II      & 0.62     & 7.8     &  30 & $-232$ & \\
9\dotfill   & 1 04 38.63  &	2 09 44.4 &     19.02 & $-$0.27 & 	O8~III((f)) & 0.18    & 2.0     &  52 & $-258$ &  \\
10\dotfill  & 1 04 43.35  &	2 10 23.4 &     18.82 & $-$0.23 & 	B1.5~Ib     & 0.21    & 2.7     &  56 & $-272$ & \\
12\dotfill  & 1 04 46.71  &	2 09 59.4 &     16.39 & 0.66 & 	G2         & 1.03     & 2.3     & 130 & $-96$ & \\
13\dotfill  & 1 04 48.98  &	2 10 24.8 &     19.80 & 0.14 & 	early A~II  & \nodata & 8.2     &  32 & $-227$ &  \\
14\dotfill  & 1 04 50.05  &	2 09 51.8 &     19.13 & $-$0.16 & 	B2~Ib      & \nodata  & 2.5     &  47 & $-232$ & \\
15\dotfill  & 1 04 43.99  &	2 07 33.0 &     15.69 & 0.91 & 	G8         & 1.00     & 2.1     & 156 & $-78$ & \hfill SK~D51\\
16\dotfill  & 1 04 47.17  &	2 07 54.2 &     18.54 & 0.39 & 	A5~Ib      & 0.82     & 5.9     &  55 & $-236$ & {\scriptsize \sc hii} \hfill SK~D22\\
17\dotfill  & 1 04 53.76  &	2 09 17.1 &     19.27 & 0.11 & 	early A~II & \nodata  & 7.1     &  43 & $-239$ & \\
18\dotfill  & 1 04 48.88  &	2 07 13.8 &     19.59 & $-$0.02 & 	B5~II      & 0.09     & 3.7     &  37 & $-224$ & \hfill SK~D28\\
19\dotfill  & 1 04 54.61  &	2 08 31.1 &     18.89 & $-$0.22 & 	composite  & 0.26     & 1.7     &  55 & $-240$ & \\
\enddata
\tablecomments{Units of right ascension are hours, minutes, and seconds, and units of declination are degrees, arcminutes, and arcseconds.}
\tablenotetext{a}{SK numbers from Sandage \& Katem (1976). The presence of nebular lines contaminating the stellar spectra is noted with {\scriptsize \sc hii}.}
\tablenotetext{b}{\hii\/ regions from Sandage (1971).}
\tablenotetext{c}{W-R star discovered by D'Odorico \& Rosa (1982).}
\end{deluxetable*}

Blue supergiant candidates were selected from $VI$ photometry measured on images
obtained with the Warsaw 1.3m telescope at Las Campanas as part of the OGLE-II
collaboration by \citet{Udalski:2001}. The spectroscopy was carried out at the
VLT UT4 (Yepun), equipped with the Focal Reducer and Low Dispersion Spectrograph
2 (FORS2), on the nights of October 26 and 27, 2003. Seeing conditions were
variable between 0\farcs8 and 1\farcs3.

\begin{figure} \epsscale{1} \plotone{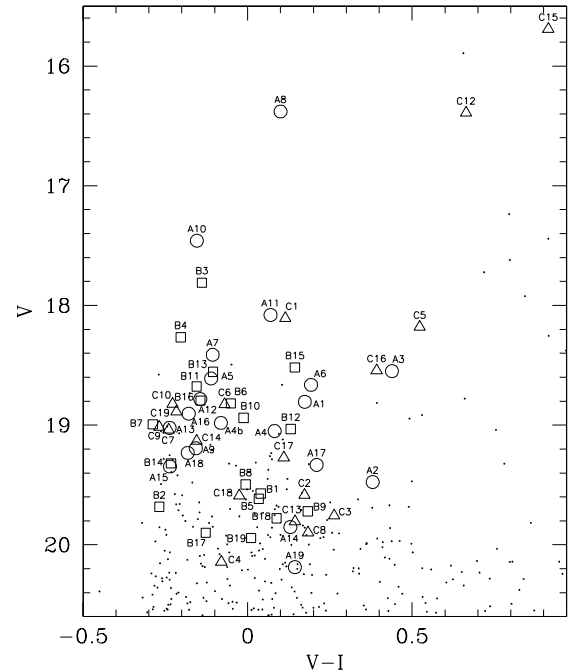} \caption{Location of the spectroscopic
targets in the $V$ vs. $V-I$ color-magnitude diagram. Different symbols are used
for stars in fields A (circles), B (squares) and C (triangles).} \label{cm}
\end{figure}

\begin{deluxetable}{cccccc}
\tabletypesize{\scriptsize}
\tablecolumns{6}
\tablewidth{0pt}
\tablecaption{Observing log\label{log}}

\tablehead{
\colhead{}	 &
\multicolumn{2}{c}{Center (J2000.0)}	&
\colhead{}	&
\colhead{}	&
\colhead{Exposure}\\
\colhead{Field}	&
\colhead{R. A.} &
\colhead{Decl.} &
\colhead{Grism}	&
\colhead{Date}	&
\colhead{Time (s)} }	
\startdata
\\[-1mm]
A\dotfill	&	01 05 01.6 & 02 09 02	&	600B	&	10-26-2003	&	3$\times$2400	\\
 		&		   &		&	1200R	&	10-26-2003	&	3$\times$2700	\\
B\dotfill	&	01 04 57.1 & 02 07 08	&	600B	&	10-27-2003	&	3$\times$2700	\\
		&		   &		&	1200R	&	10-27-2003	&	3$\times$2700	\\
C\dotfill	&	01 04 38.9 & 02 09 07	&	600B	&	10-26-2003	&	2$\times$1800, 1$\times$700	\\
		&		   &		&	1200R	&	10-27-2003	&	2$\times$1200, 1$\times$1800	\\
\enddata
\end{deluxetable}

We acquired spectra in three 6\farcm8 $\times$ 6\farcm8 FORS2 fields in the
movable slitlets (MOS) mode (19 slitlets per field, each 1~arcsec wide and
extending $\sim$ 21 arcsec in the spatial direction). We included as many of the
brightest blue stars ($V<20$, $V-I<0.4$) found in the three fields of IC~1613 as
possible. A handful of redder or fainter stars were also observed, in order to
fill up the MOS setup. A finding chart is shown in Fig.~\ref{chart}, while
target coordinates, photometric measurements and other derived parameters are
presented in Table~\ref{catalog}. Fig.~\ref{cm} shows the location of the
spectroscopic targets in the color-magnitude diagram. Throughout this paper we
adopt the convention of appending the slit number (1 to 19) to the field
identification (A, B or C) to label our spectroscopic targets. Our field A is
centered on the NE quadrant of IC~1613, where the majority of the young stars
and \hii\/ regions are located. Field B covers most of the central bar-like
structure that runs from the SE to the NW, while field C extends to the west of
these two main structures. Two stars were included in the 4th slitlet of field
A, and are labeled as A4 and A4b. Star C11 was too close to the slitlet edge to
provide a meaningful extraction, and therefore it does not appear in
Table~\ref{catalog}.

The spectroscopy was carried out using two different grisms: the 600B to cover
the blue wavelengths (in particular the 3600-5000~\AA\/ range, which is critical
for the spectral classification) at 5~\AA\/ resolution, and the 1200R to cover
the region around the H$\alpha$ line at 3~\AA\/ resolution. The integration
times varied among the different fields and setups, and are summarized in
Table~\ref{log}. As field C contained our lowest-priority targets, the exposure
times were shorter than in the remaining two fields. The airmass of IC~1613 was
also larger during the spectroscopy of field C, exceeding a value of 1.6. For
fields A and B the airmass varied between 1.1 and 1.5.

We have used standard long-slit spectroscopy tasks within {\sc
iraf}\footnote{{\sc iraf} is distributed by the National Optical Astronomy
Observatories, which are operated by the Association of Universities for
Research in Astronomy, Inc., under cooperative agreement with the National
Science Foundation.} for bias and flat-field corrections, wavelength calibration
and extraction of the spectra. Cosmic rays were removed with the {\sc
l.a.cosmic} routines by \citet{van-Dokkum:2001}. The mean signal-to-noise ratio
(S/N) of the resulting averaged, normalized spectra is reported in column~9 of
Table~\ref{catalog}. The heliocentric radial velocities measured from the
spectral lines (column 10) are in all cases but 3 in good agreement with the
{\sc h~i} systemic velocity of $-234\pm1$ km\,s$^{-1}$ (\citealt{Lu:1993}).
Excluding stars C12 and C15 (2 foreground G stars) and C5 (presumably a Galactic
halo A star), we obtain a mean radial velocity of $-241\pm18$ km\,s$^{-1}$. With
an uncertainty of $\sim$10~km\,s$^{-1}$ on the radial velocities, and a maximum
rotational velocity of $\sim$20~km\,s$^{-1}$ from \hi\/ observations
(\citealt{Lake:1989}, \citealt{Hoffman:1996}), we are finding no hints of
galactic rotation from our stellar data.

Cross identification with the bright star catalog of \citet{Sandage:1976} is
given in column 11 of Table~\ref{catalog}. We also indicate the presence of
spatially extended nebular emission superposed on the stellar spectra. In some
cases this contamination strongly affects the stellar lines, in particular those
of the Balmer series.

\section{Spectral catalog}

\begin{figure*} \epsscale{0.9} \plotone{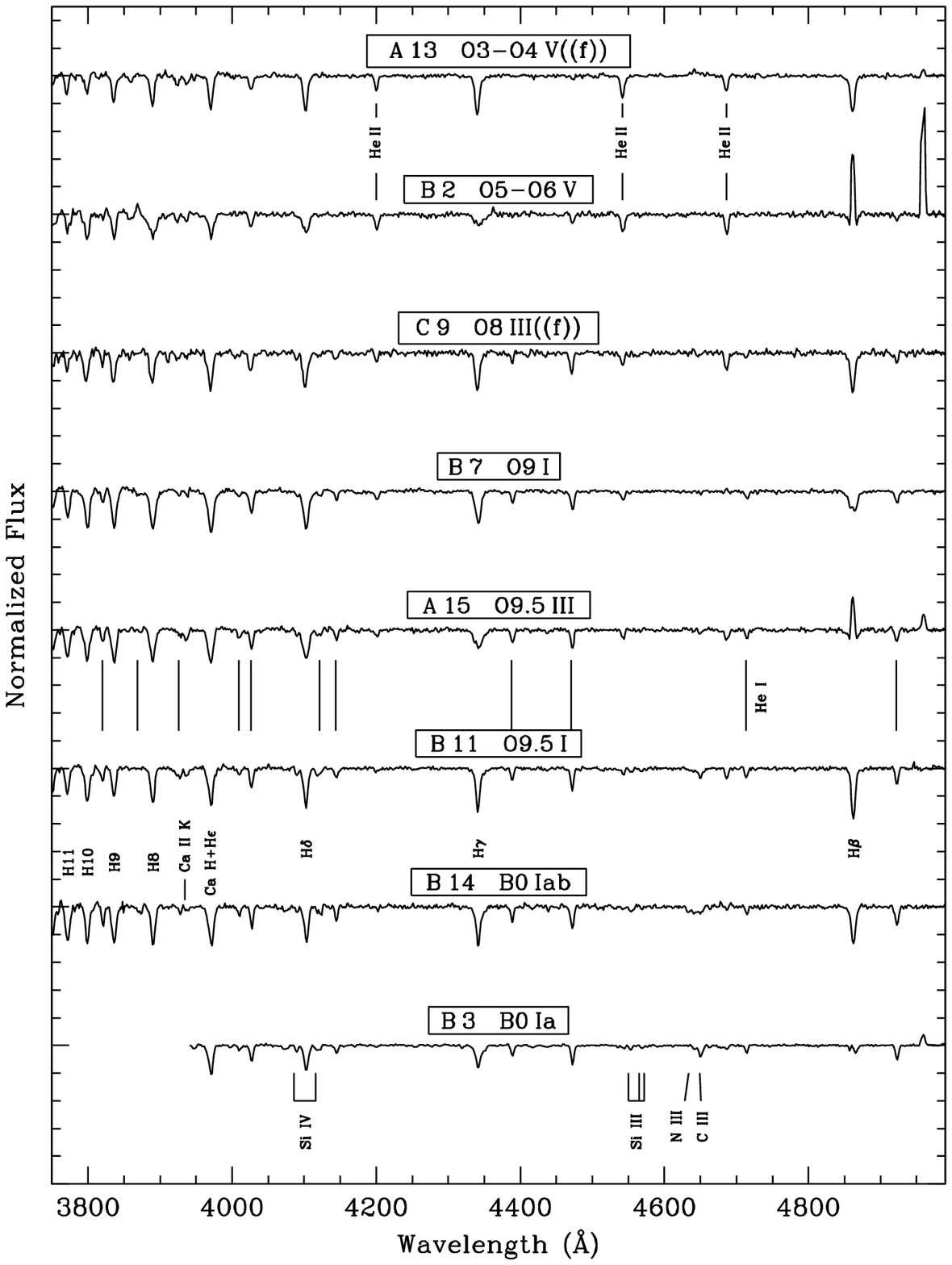} \caption{Normalized spectra of the O
and B0 stars. The spectral features identified are \heii\,\llin4200, 4542, 4686
{\em (below A13)}; \hei\,\llin3820, 3867-3872, 3926, 4009, 4026, 4121, 4144,
4388, 4471, 4713, 4922 {\em (below A15)}; Balmer lines from H$\beta$ to H11, and
the \caii\,K (\lin3933) and H (\lin3968) lines {\em (below B11)};
\siiv\,\llin4089, 4116, \siiii\,\llin4553, 4568, 4575, \niii\,\llin4634,
4640-4642, \ciii\,\lin4650 {\em (below B3)}. Here, and through Fig.~\ref{spectra7},
ordinate tick marks are drawn at intervals of 0.2 continuum flux units.}
\label{spectra1} \end{figure*}

\begin{figure*} \epsscale{0.9} \plotone{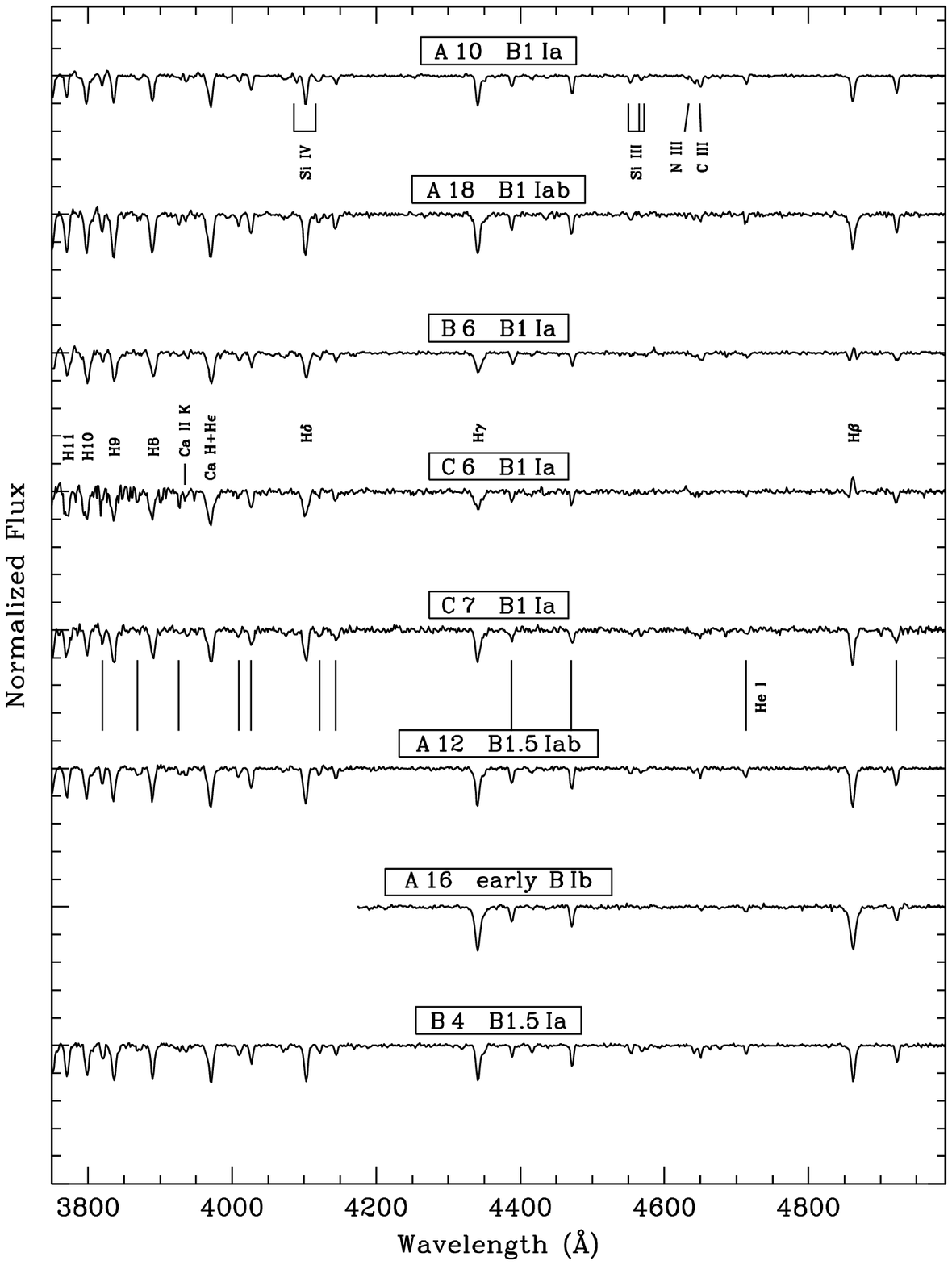} \caption{Normalized spectra of
early-B stars (B1--B1.5). The spectral features identified are \siiv\,\llin4089,
4116, \siiii\,\llin4553, 4568, 4575, \niii\,\llin4634, 4640-4642,
\ciii\,\lin4650 {\em (below A10)}; Balmer lines from H$\beta$ to H11, and the
\caii\,K (\lin3933) and H (\lin3968) lines {\em (below B6)}; \hei\,\llin3820,
3867-3872, 3926, 4009, 4026, 4121, 4144, 4388, 4471, 4713, 4922 {\em (below
C7)}. } \label{spectra2} \end{figure*}

\begin{figure*} \epsscale{0.9} \plotone{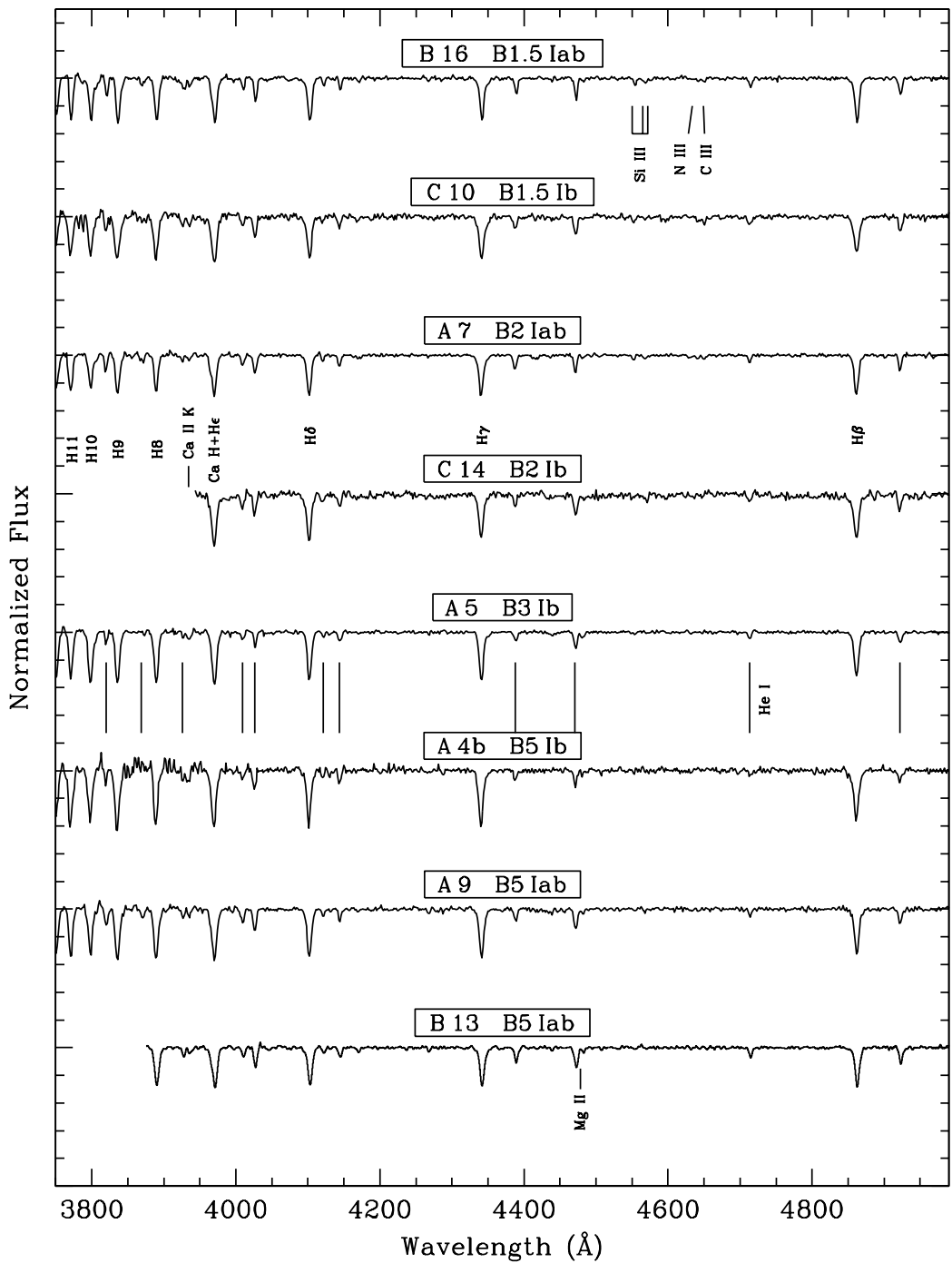} \caption{Normalized spectra of
early- to  mid-B stars (B1.5--B5). The spectral features identified are
\siiii\,\llin4553, 4568, 4575, \niii\,\llin4634, 4640-4642, \ciii\,\lin4650 {\em
(below B16)}; Balmer lines from H$\beta$ to H11, and the \caii\,K (\lin3933) and
H (\lin3968) lines {\em (below A7)}; \hei\,\llin3820, 3867-3872, 3926, 4009,
4026, 4121, 4144, 4388, 4471, 4713, 4922 {\em (below A5)} and \mgii\,\lin4481
{\em (below B13)}. } \label{spectra3} \end{figure*}

\begin{figure*} \epsscale{0.9} \plotone{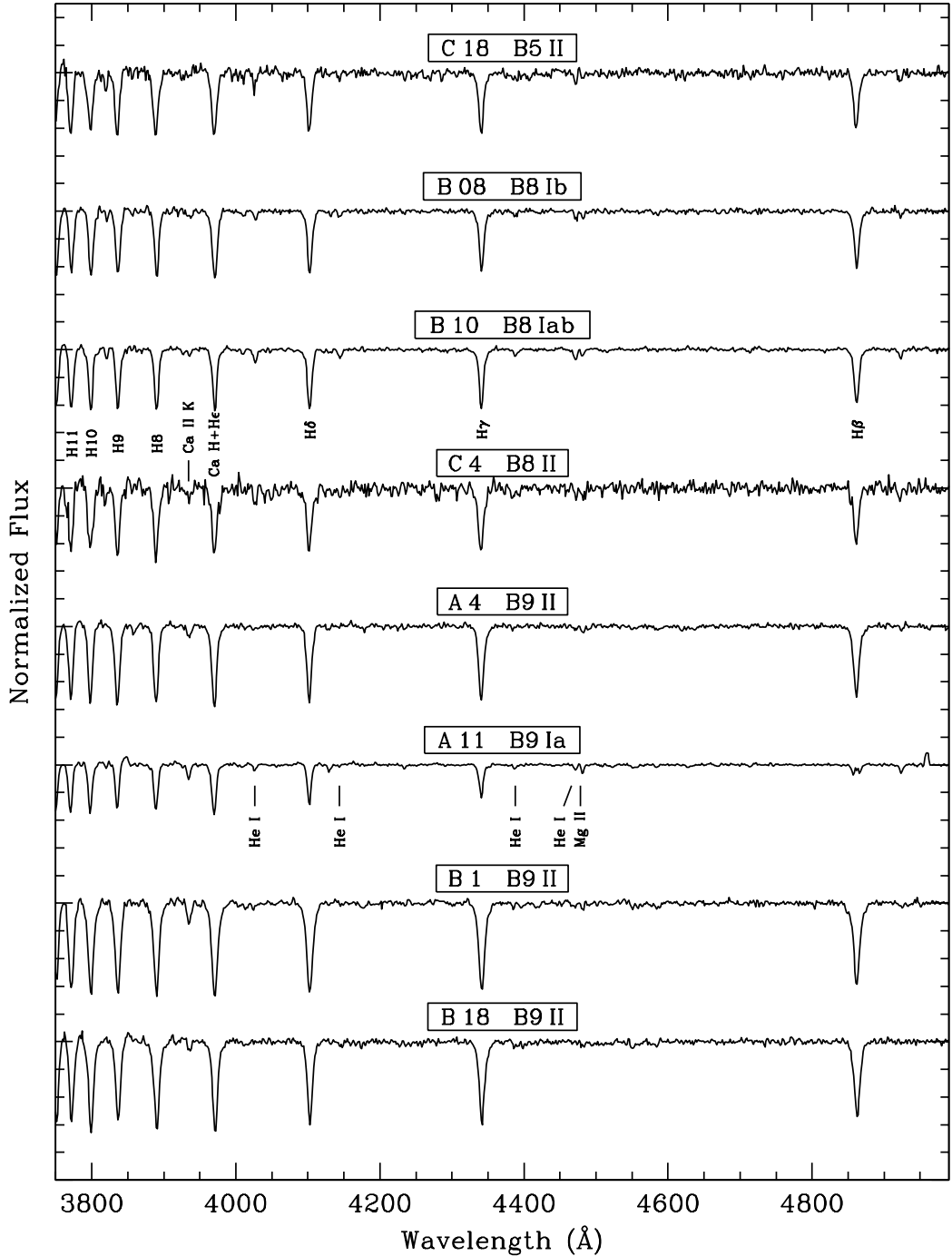} \caption{Normalized spectra of mid-
to  late-B stars (B5--B9). The spectral features identified are Balmer lines
from H$\beta$ to H11, and the \caii\,K (\lin3933) and H (\lin3968) lines {\em
(below B10)}; \hei\,\llin4026, 4121, 4388, 4471 and \mgii\,\lin4481 {\em (below
A11)}. } \label{spectra4} \end{figure*}

\begin{figure*} \epsscale{0.9} \plotone{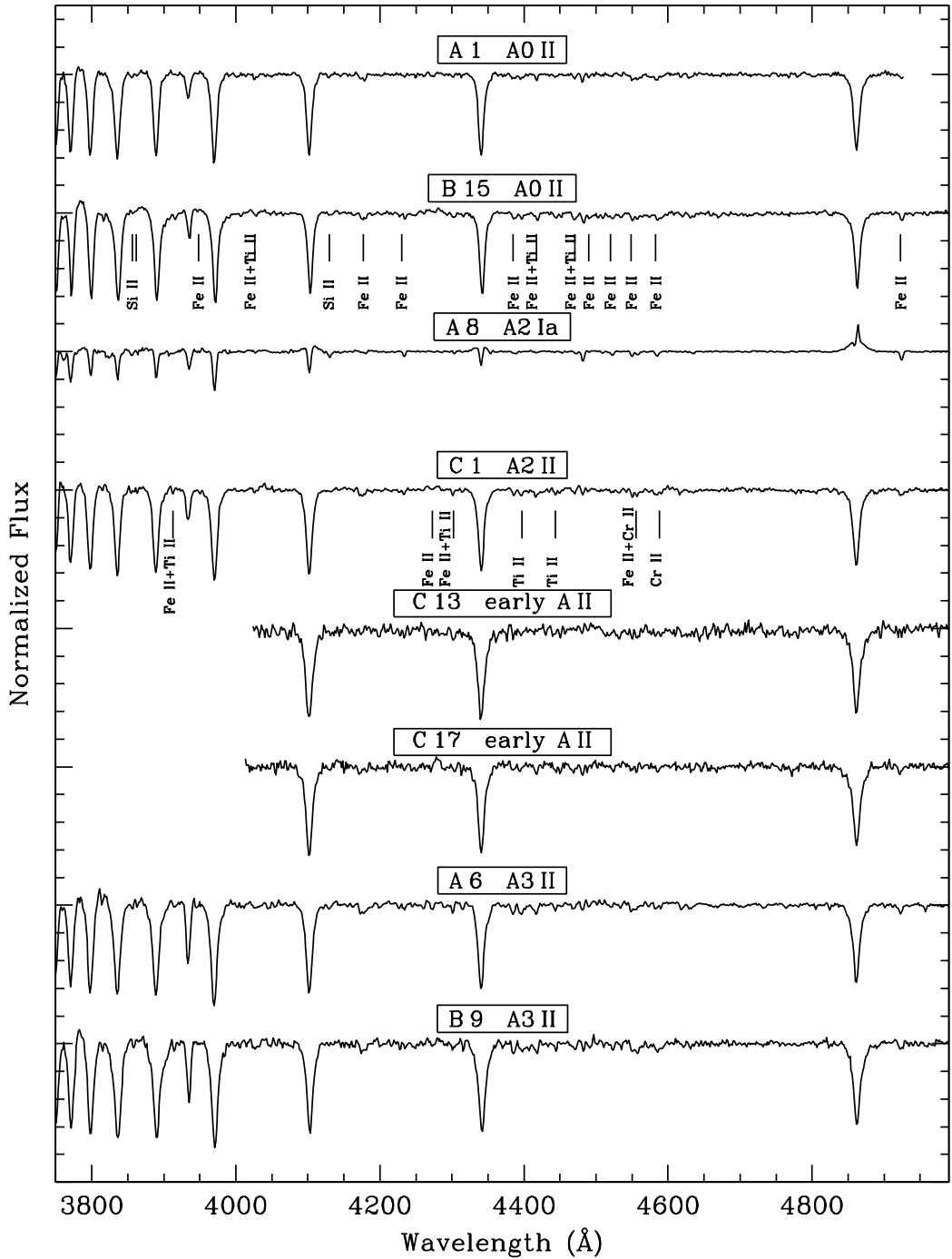} \caption{Normalized spectra of
early-A stars (A0--A3). The spectral features identified are
\siii\,\llin3856-3862, 4128-4132, \feii\,\llin3945, 4024, 4173-4178, 4233, 4385,
4417, 4473, 4489-4491, 4549, 4583, 4923, and \tiii\,\llin4028, 4418, 4469-4471
{\em (below B15)}; \feii\,\llin3914, 4273, 4303, 4556, \tiii\,\llin3913, 4300,
4395-4399, 4444, and \crii\,\llin4588, 4559 {\em (below C1)}. } \label{spectra5}
\end{figure*}

\begin{figure*} \epsscale{0.9} \plotone{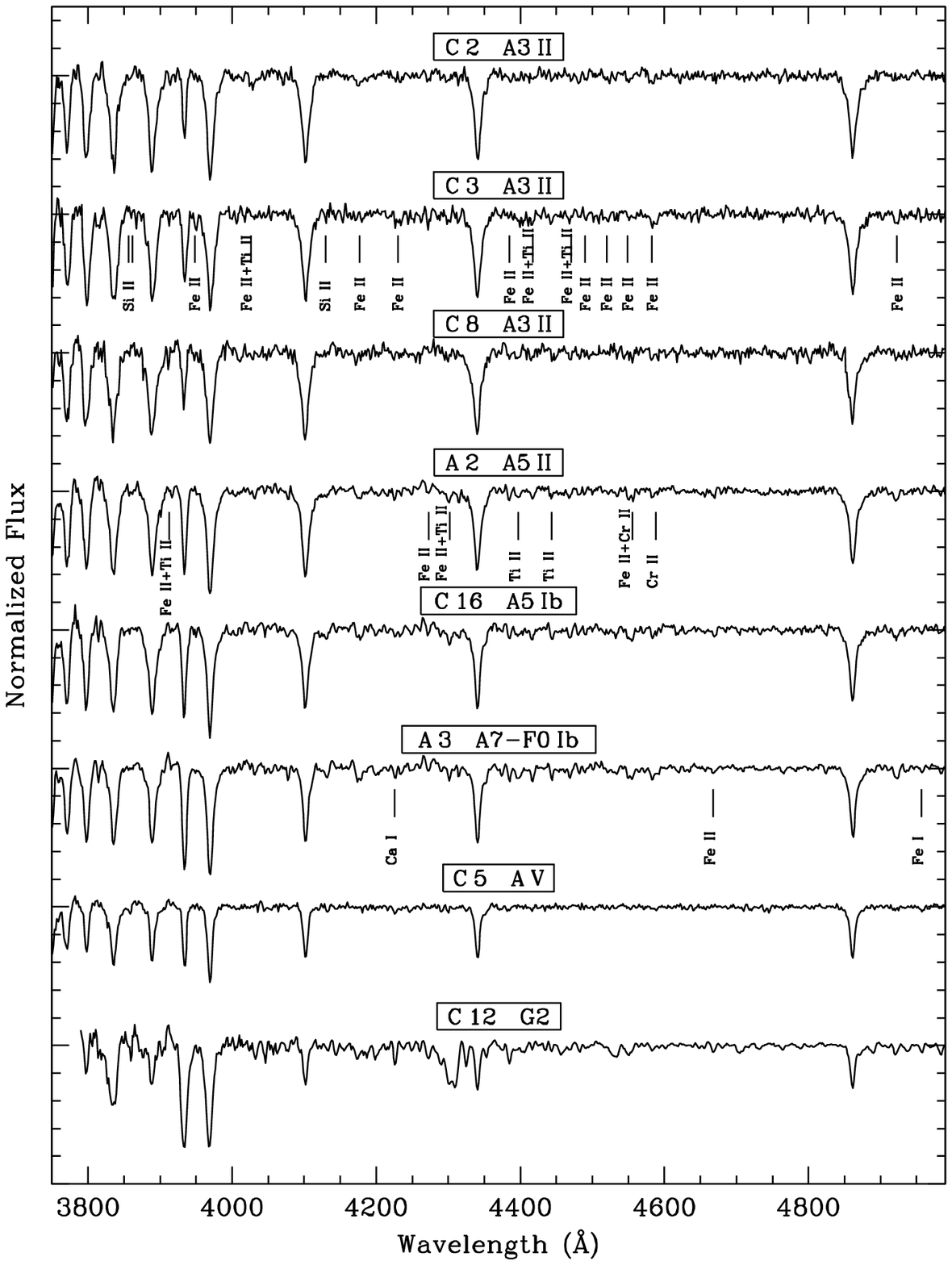} \caption{Normalized spectra of mid-A
to G stars (A3--G2). The last two stars shown, C5 and C12, are foreground
Galactic objects. The spectral features identified are \siii\,\llin3856-3862,
4128-4132, \feii\,\llin3945, 4024, 4173-4178, 4233, 4385, 4417, 4473, 4489-4491,
4549, 4583, 4923, and \tiii\,\llin4028, 4418, 4469-4471 {\em (below C3)};
\feii\,\llin3914, 4273, 4303, 4556, \tiii\,\llin3913, 4300, 4395-4399, 4444, and
\crii\,\llin4588, 4559 {\em (below A2)}; \feii\,\llin4666-4669, \fei\,\lin4957,
and \cai\,\lin4226 {\em (below A3)}. } \label{spectra6} \end{figure*}

\begin{figure*} \epsscale{0.9} \plotone{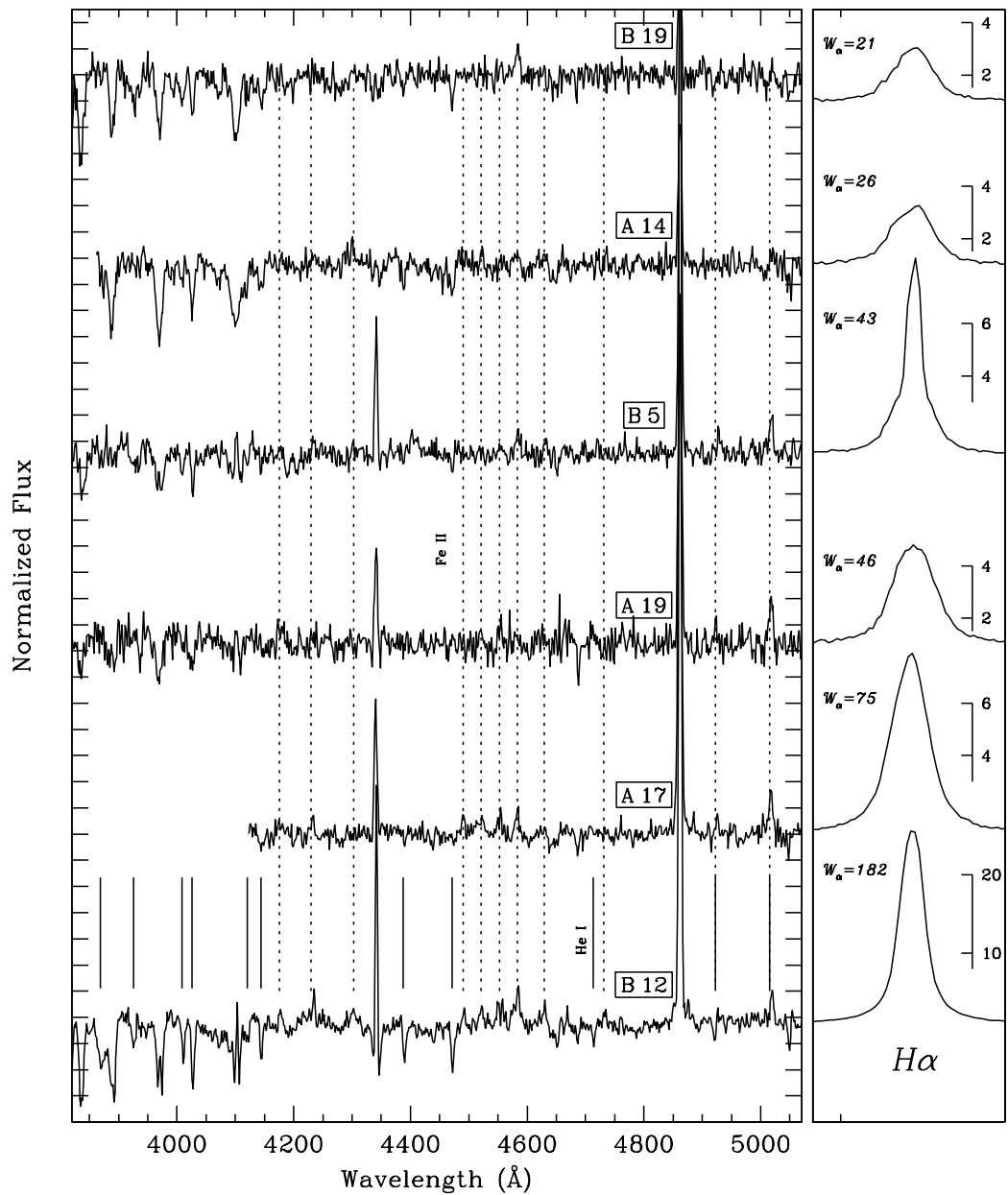} \caption{Normalized spectra of the
Be stars, in order of increasing strength of the H$\alpha$ emission (the
H$\beta$ line is clipped for clarity). The dotted lines mark the wavelengths of
the \feii\, lines \llin4173-4178, 4233, 4303, 4489-4491, 4520-4523, 4549-4556,
4583, 4629, 4731, 4922, 5015. The wavelengths of the following neutral helium
lines are indicated at the bottom: \hei\llin3867-3872, 3926, 4009, 4026, 4121,
4144, 4388, 4471, 4713, 4922, 5015. The panel on the right displays the profile
of the H$\alpha$ line, using different vertical stretches. The equivalent width
of the line, $W_\alpha$, is indicated to the left of each profile. The spectrum
of star B5 is affected by nebular emission. } \label{spectra7} \end{figure*}

\begin{figure*} \epsscale{1} \plotone{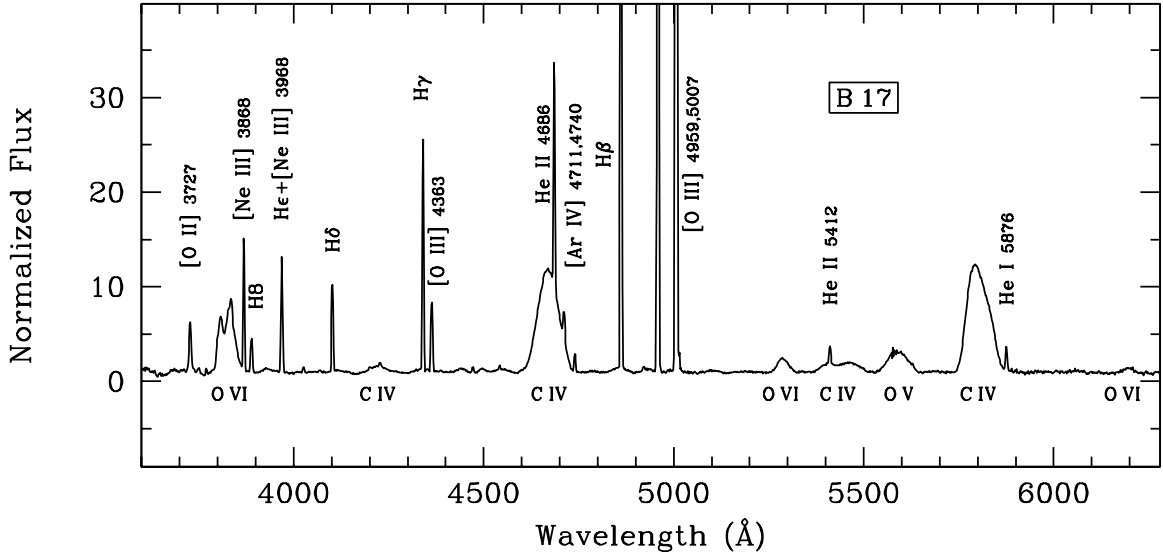} \caption{Normalized spectrum of the
WO3 star B17. Broad stellar lines are identified at the bottom: \ov\,\lin5590,
\ovi\,\llin3811-3834, 5290, 6200 and \civ\,\llin4229, 4658-4686, 5411-5470,
5801-5812. A number of nebular emission lines are also marked in the top portion
of the figure. } \label{spectra8} \end{figure*}

\begin{figure*} \epsscale{0.9} \plotone{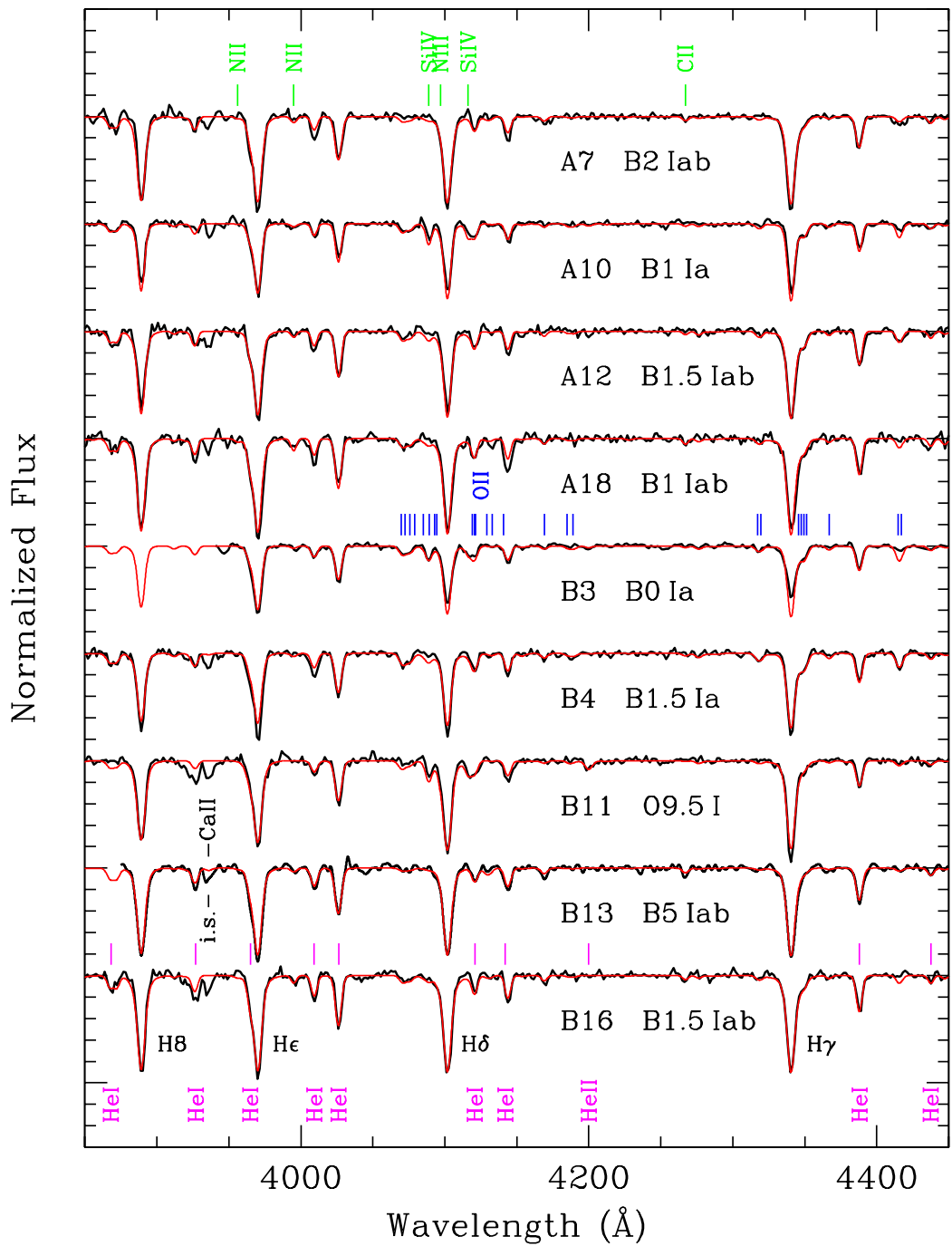} \caption{Comparison between the
observed spectra (thick dark lines) and the adopted {\sc fastwind} models (red
lines) in the 3850--4450~\AA\/ range. Identifications for the H, He and some of
the metal lines are provided. Here, and in Fig.~\ref{models2}, ordinate tick marks are drawn at intervals of 0.2 continuum flux units.
} \label{models1} \end{figure*}

\begin{figure*} \epsscale{0.9} \plotone{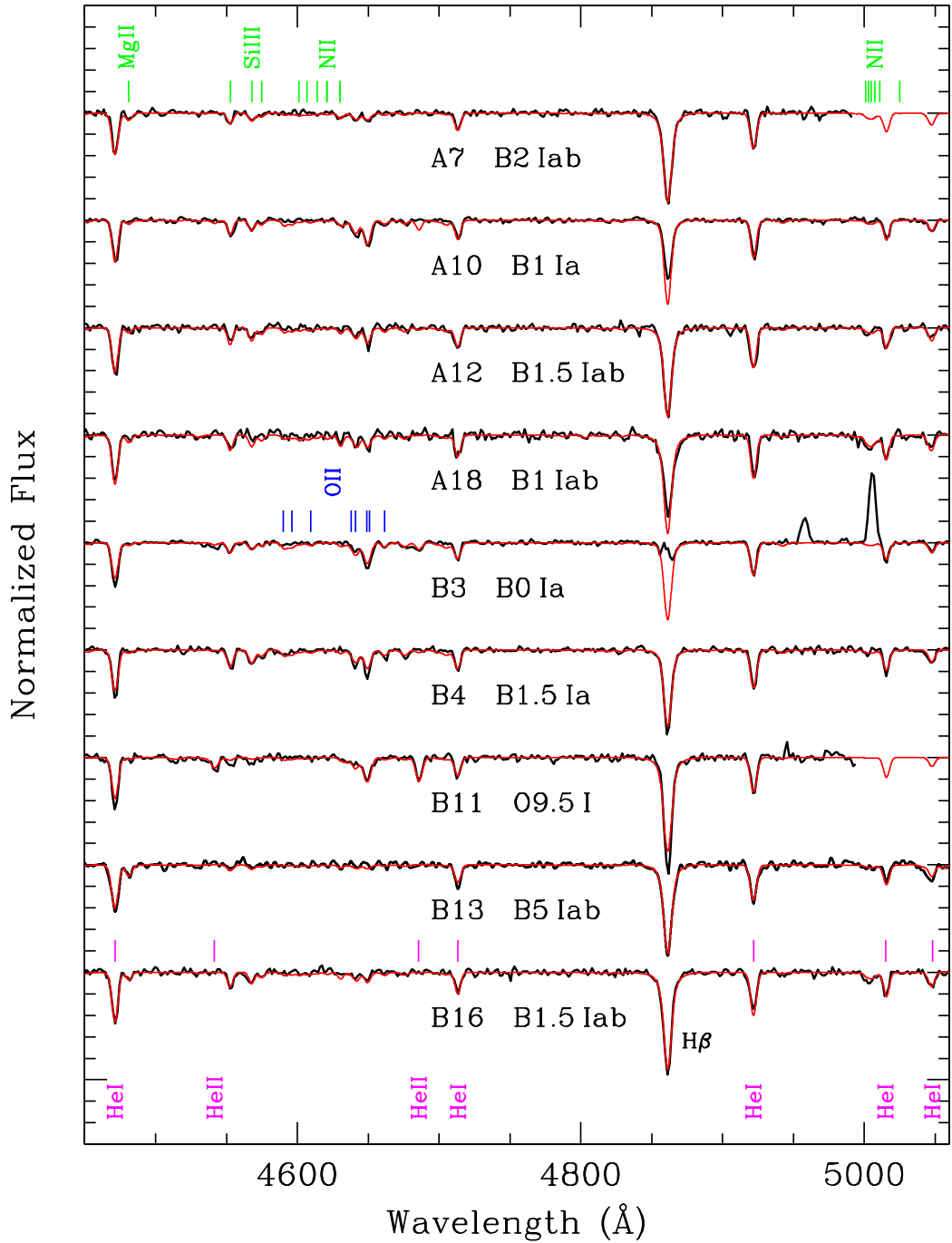} \caption{Same as
Fig.~\ref{models1}, for the 4450--5065~\AA\/ range. } \label{models2}
\end{figure*}

\begin{figure*} \epsscale{0.9} \plotone{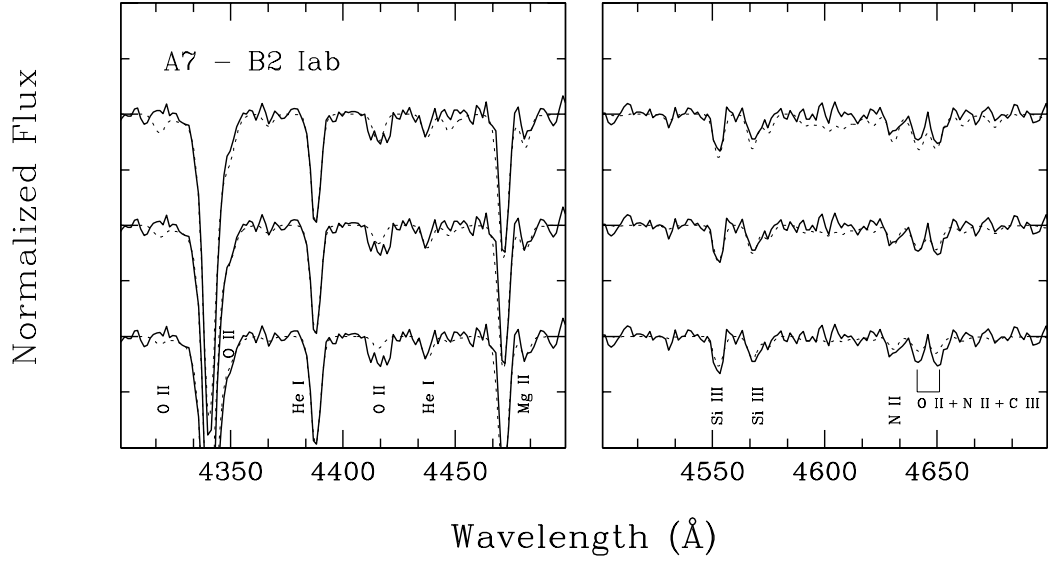} \caption{Comparison between 
the observed spectrum of the B2~Iab star A7 (solid lines) with {\sc fastwind} models
calculated by varying the metal abundances by $\pm0.2$ dex (dotted lines, top and bottom spectra, 
respectively) relative to the adopted metallicity (dotted line, central spectra).
The spectral features identified are O\,{\sc ii}\,\llin4317-4319, 4346-4351, 4415-4417; 
\hei\,\llin4388, 4438; \mgii\,\lin4481; \siiii\,\llin4553, 4568, 4575; 
\nii\,\lin4630, and the O\,{\sc ii}~+~\nii~+~\ciii\/ blend at \lin4650.} \label{models4}
\end{figure*}

\begin{figure*} \epsscale{1} \plotone{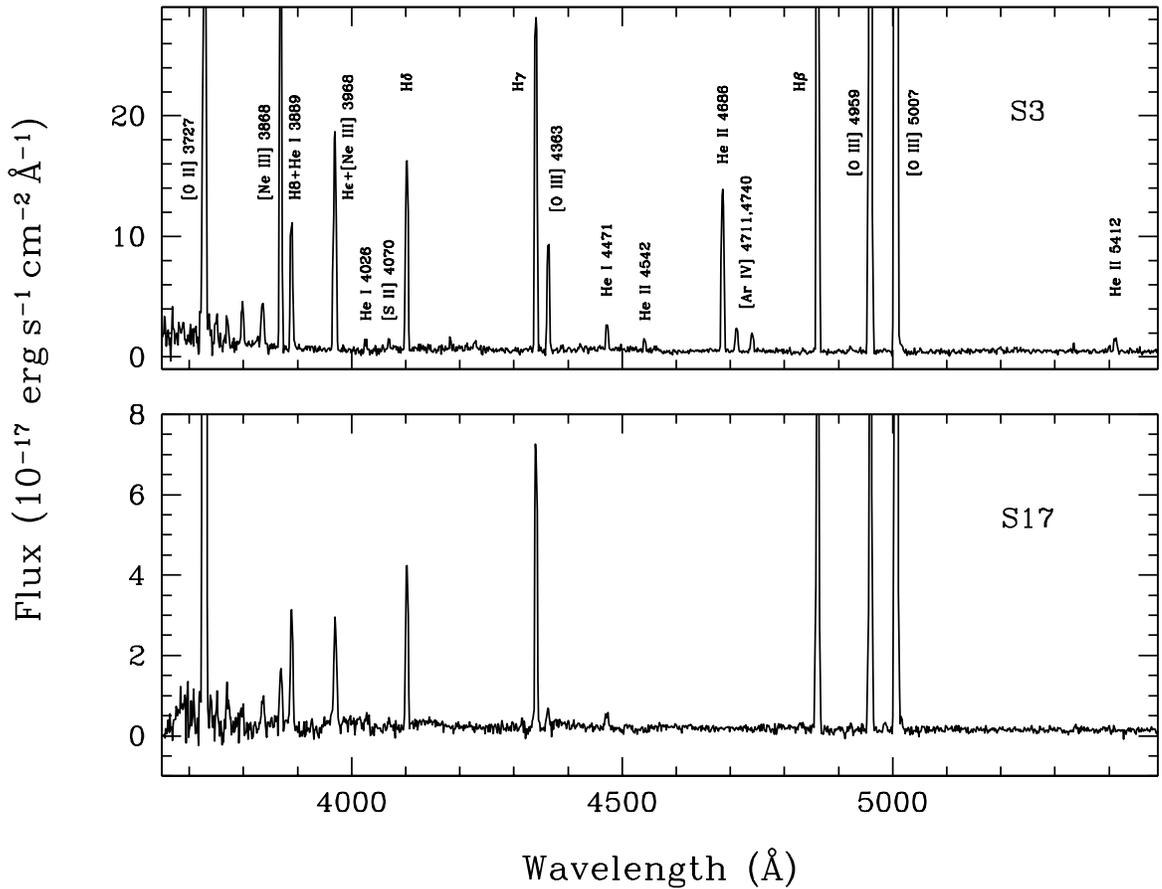} \caption{Spectra of the \hii\/
regions S3 (top) and S17 (bottom). The main nebular lines are identified.}
\label{spectrahii} \end{figure*}

The philosophy behind the classification of extragalactic early-type stars has
been discussed at length in our previous works on WLM (\citealt{Bresolin:2006})
and NGC~3109 (\citealt{Evans:2007}). We recall here that for B stars we follow
\citet{Lennon:1997}, and for A supergiants we adopt the criteria from
\citet{Evans:2003} and \citet{Evans:2004}. The classification of O stars is
based upon \citet{Walborn:1971}, \citet{Walborn:1990} and \citet{Walborn:2000}.
Luminosity classes for B and A stars are determined from the equivalent width of
H$\gamma$ ($W_\gamma$), following the criteria by \citet{Azzopardi:1987}.

The spectral classification of the 57 stars included in this work are given in
column 6 of Table~\ref{catalog}, together with the
\caii~K/(\caii~H~+~H$\epsilon$) ratio used for classifying A stars (column 7)
and $W_\gamma$ (column 8). The blue region (3750~\AA\/ to 5000~\AA) of the
normalized stellar spectra is shown in Fig.~\ref{spectra1}--\ref{spectra6}, in
order of decreasing temperature (spectral type O to G). Emission-line stars are
shown in Fig.~\ref{spectra7} and \ref{spectra8}. In these figures we identify
the main H, He and metal lines that are used in the classification procedure.

\subsection{Comments on selected stars} We provide here brief comments on some
peculiar or outstanding  stars presented in Table~\ref{catalog} and shown in
Figures~\ref{spectra1}-\ref{spectra8}. Absolute magnitudes are derived from a
distance modulus (m$-$M)$_0$~=~$24.291 \pm 0.035$ and E(B$-$V)~=~$0.090 \pm 0.019$
(\citealt{Pietrzynski:2006}).

{\rm A8}\,{\em (A2~Ia).}---This is the visually brightest star in IC~1613 at
$M_V=-8.2$. \citet{Sandage:1976} suggested that this star (their A43) is a
foreground field star, but its radial velocity was found to agree with the
systemic velocity by \citet{Humphreys:1980}, who assigned a A0~Ia spectral type.
The H$\alpha$ (not shown) and H$\beta$ lines are in emission, with P Cygni
profiles. The other non-Be stars with strong wind emission in H$\alpha$ are the 
B1 supergiants B6 and C6.

{\rm A10}\,{\em (B1~Ia).}---Star 22A in the list of \citet{Sandage:1976}, it was
found by these authors to be the brightest blue star in IC~1613 ($M_V=-7.1$).
\citet{Humphreys:1980} classified this star as B2-B3~I. We assign the earlier B1
type, from the detection of \siiv\llin4089, 4116 (the latter line blended with
\hei\lin4121), and the lack of \heii\/ lines.

{\rm A13}\,{\em (O3-O4~V((f))).}---The earliest star in our sample shows no
evidence of \hei\/ in its spectrum, except perhaps a weak \hei\lin4471 line.
Since nebular lines (e.g.~\oiii\lin5007) are present in our spectrum, this lack
of \hei\/ lines could be due to line in-filling. The strong \heii\lin4686
absorption and the weak   {\sc n\,iii}\llin4634-4640-4642 emission lines lead to
the ((f)) classification. A weak {\sc n\,iv}\lin4058 in emission is detected.
The absolute magnitude of this star, $M_V=-5.5$, is consistent with the dwarf
classification (\citealt{Conti:1988}).

{\rm B2}\,{\em (O5-O6\,V).}---This O star displays \hei\/ lines in its spectrum,
with \heii\lin4200~$>$~\hei\lin4026 (the two lines are of equal strength at
class O6). Its value of $\log W^\prime = \log W($\hei\lin4471$)-\log
W($\heii\lin4542$)=-0.42$ corresponds to an O5.5 type (\citealt{Conti:1988}).
This star is the likely ionizing source of the surrounding \hii\/ region S17
(\citealt{Sandage:1971}), whose chemical abundance is derived in
Section~\ref{nebular}. The contamination by the nebular lines is causing
in-filling of the \hei\/ lines, therefore the spectral type can be somewhat
later than O5-O6. From the H$\alpha$ luminosity of S17 published by
\citet{Hodge:1990}, log~L(H$\alpha$)~=~36.96 erg\,s$^{-1}$, we derive the number
of Lyman continuum ionizing photons log~Q(H$^0$)~=~48.82. This corresponds to
the ionizing flux output of an O6.5\,V star (\citealt{Martins:2005}).

{\rm B3}\,{\em (B0~Ia).}---The third brightest blue supergiant ($M_V=-6.8$) is
star B42 in the list by \citet{Sandage:1976}. Our B0 classification derives from
the presence of \heii\llin4542, 4686 in its spectrum, with
\heii\lin4542~$<$~\siiii\lin4553 (the two lines are of equal strength at O9.7).
The stellar spectrum is heavily contaminated by nebular lines (see, for example,
the filled-in H$\beta$ line and \oiii\lin5007 in emission).
\citet{Humphreys:1980} obtained a spectrum of this star, and assigned a B0
classification based on two-color photometry. This is the only star we have in
common with \citet{Lozinskaya:2002}, who classified it (their star I.1) as an O supergiant.

{\rm B17}\,{\em (WO3).}---The Wolf-Rayet features in the bright star embedded in
the \hii\/ region S3 were studied for the first time by \citet[hence the name
DR1]{Dodorico:1982} and \citet{Davidson:1982}. Spectra of this star have been
discussed, among others, by \citet{Armandroff:1991}, who classified it as WC4-5,
\citet[WO4]{Garnett:1991} and \citet[WO3]{Kingsburgh:1995}, whose classification
we adopt. The spectrum (Fig.~\ref{spectra8})  is characterized by
high-excitation nebular lines (including \heii\llin4542, 4686, 5412) superposed
on broad stellar lines from \civ, \ov\/ and \ovi.

{\rm C5}\,{\em (A~V).}---This star has a highly discrepant radial velocity,
$V_{\rm Helio}=+17$~km~s$^{-1}$, and its spectrum is characterized by narrow
Balmer lines and very weak metal lines, with the exception of \caii~K. A halo
dwarf A-type star or a lower-gravity, evolutionarily more evolved field
horizontal branch star would both match these properties. Disentangling these
two possibilities from linewidth measurements, even at higher resolution, is
unfeasible for the coolest ($B-V>0.2$) A stars\footnote{We follow the
terminology of \citet{Wilhelm:1999a} and other works dealing with halo stars by
including in the A-type class stars as cool as \teff~=~6000, i.e.~what would be
traditionally called F-type stars.} (\citealt{Arnold:1992}). However, {\em UBV}
photometry can be used to constrain \teff\/ and surface gravity, and we note
that in the $(U-B)_0$~vs.~$(B-V)_0$ color-color diagram shown by
\citet{Wilhelm:1999a} in their extensive study of A-type stars in the halo star
C5 would fall in the A~V domain. We adopted $U-B=-0.18$ and $B-V=0.37$
(A.~Herrero, private communication), and a Galactic reddening  towards IC~1613
$E(B-V)=0.025$ (\citealt{Schlegel:1998}). We have used the model grid by
\citet{Wilhelm:1999b} to obtain approximate values of the stellar parameters
from the {\em UBV} colors and the equivalent width of the \caii~K line. Our
results, \teff~=~6750~$K$, \logg~=~4.0, and [Fe/H]~=~$-2.0$, confirm that C5 is
a dwarf star. From an absolute magnitude $M_V=3.6$, corresponding to the star's
$(B-V)_0$ color (\citealt{Schmidt-Kaler:1982}), we derive a distance of 8.1 kpc.
We note that the \caii~K/(\caii~H~+~H$\epsilon$) criterion provides an A5
classification.

{\em Be stars}---The group of 6 stars in Fig.~\ref{spectra7} displays similar
characteristics, with Balmer lines, as well as several \feii\/ lines, in
emission. While H$\alpha$ and H$\beta$ are in emission for all stars, H$\gamma$
progresses from a filled-in profile to an emission line as EW(H$\alpha$)
increases. The strength of the \feii\/ emission appears to be correlated with
that of the H lines. The \hei\lin5876 is seen in emission in all stars. The
simultaneous presence of H, He and iron emission lines in B stars is consistent
with the Group~I Be classification of \citet{Jaschek:1980}. The definition of
the B sub-class is made highly uncertain by the broadening of the lines (due to
high rotational velocities) and the superposition of the \feii\llin4549-4556,
4583 emission lines, originating in the surrounding shells, on top of the
photospheric \siii\llin4553, 4568, 4575 lines. However, Be stars belonging to
Group~I are mostly of early-B type (B0-B3), and we detect a broad feature
corresponding to the CNO blend at $\sim4650$~\AA, found in early-B stars. We
also detect \heii\lin4686 in a few cases (e.g.~A19, A17), pointing to an
earlier, Oe star classification (\citealt{Conti:1974}). Absolute visual
magnitudes range between $M_V=-5.6$ (B12) and $M_V=-4.4$ (A19). These values are
comparable to those found for the brightest Be stars in the SMC
(\citealt{Martayan:2007}). The large number of Be stars found in our study tends
to confirm the trend of increasing frequency of Be stars with decreasing
metallicity  (\citealt{Maeder:1999,Wisniewski:2006}).

{\em G stars}---Both C12 (type G2) and C15 (type G8, not shown in the spectral
catalog) have a radial velocity which is discrepant with respect to the systemic
velocity of IC~1613, as well as a bright apparent magnitude ($V$~=~16.39 and
15.69, respectively), indicating that they are foreground Milky Way stars. The
classification is based on the strength of the G-band and \fei\lin4325 relative
to H$\gamma$ (\citealt{Evans:2004}). The photometric parameters for these stars
violate the criteria we use to select blue supergiants in external galaxies.
These objects were included in the spectroscopic sample in order to fill
otherwise unused slitlets of our MOS setup.

\section{Stellar abundances: B supergiants}

For the 9 early-B supergiants with the highest signal-to-noise ratio we have
derived stellar parameters and metal abundances by comparing our observations
with model spectra obtained with the {\sc fastwind} code
(\citealt{Santolaya-Rey:1997}, \citealt{Puls:2005}). Our procedure has been
explained in detail by \citet{Urbaneja:2003,Urbaneja:2005}, and has been
recently used for the analysis of B supergiants in WLM (\citealt{Bresolin:2006})
and NGC~3109 (\citealt{Evans:2007}). As already mentioned in these papers, our
best-fit models are those that provide the overall best match to the ensemble of
metal lines present in the spectra, while some of the individual lines might be
still not well reproduced. At the low abundances encountered in the dwarf
irregular galaxies that we have so far analyzed [12~+~log(O/H)~$<$~8.0], the low
spectral resolution and the limited signal-to-noise ratio that are available to
us conspire to increase the uncertainty in the metallicity measurements, which
for a single chemical element in a given star is on the order of 0.2-0.25 dex.
However, combining the information obtained from several stars, we can still
obtain a meaningful picture of the content of the most abundant elements, in
particular oxygen, in galaxies at relatively large distances, and verify, within
the above-mentioned limits of precision, the presence of chemical
inhomogeneities  and the agreement with chemical abundances obtained with
alternative means, in particular from the emission lines of \hii\/ regions
(\citealt{Bresolin:2006}).

\hoffset=-20 mm
\begin{deluxetable*}{cccccccccc}
\tabletypesize{\scriptsize}
\tablecolumns{10}
\tablewidth{0pt}
\tablecaption{Physical parameters of B supergiants\label{modelparameters}}

\tablehead{
\colhead{\phantom{}Properties\phantom{abcdefg}}	 &
\colhead{A7}     	&
\colhead{A10}           &
\colhead{A12}           &
\colhead{A18}		&
\colhead{B3}		&
\colhead{B4}		&
\colhead{B11}		&
\colhead{B13}		&
\colhead{B16}		}
\startdata
\\[-1mm]
Spectral type \dotfill	 &  B2~Iab &  B1~Ia & B1.5~Iab & B1~Iab & B0~Ia & B1.5~Ia & O9.5~Iab & B5~Iab & B1.5~Iab \\
\teff\/ (K)\tablenotemark{a}\dotfill	& 19500 & 25000 & 23000 & 21000 & 24500 & 22500 & 30000 & 17000 & 21000  \\
log $g$ (cgs)\tablenotemark{b}\dotfill	& 2.55 & 2.70 & 2.75 & 2.70 & 2.65 & 2.60 & 3.20 & 2.30 & 2.75    \\
$R/R_\odot$\dotfill	& $30\pm3$ & $38\pm3$ & $24\pm2$ & $19\pm2$ & $32\pm3$ & $27\pm2$ & $19\pm1$ & $31\pm3$ & $25\pm3$ \\
B.C.\dotfill		& $-1.89$ & $-2.47$ &  $-2.32$  & $-2.08$ & $-2.42$ & $-2.19$ & $-2.19$ & $-1.57$ & $-2.09$          \\
M$_{\rm bol}$\dotfill	& $-7.93$ & $-9.55$ & $-8.13$ & $-7.14$ & $-9.10$ & $-8.26$ & $-8.80$ & $-7.40$ & $-7.86$   \\
log $L/L_\odot$\dotfill	& $5.07\pm0.18$ & $5.71\pm0.14$ & $5.15\pm0.16$  & $4.75\pm0.18$ & $5.53\pm0.14$ & $5.20\pm0.15$ & $5.42\pm0.11$ &
$4.86\pm0.20$ & $5.04\pm0.16$    \\
$M_{\rm spec}/M_\odot$\dotfill  & $12\pm5$ & $27\pm10$ & $11\pm5$ & $7\pm3$ & $19\pm7$ & $11\pm4$ & $21\pm7$ & $7\pm3$ & $13\pm6$       \\
$E(B-V)$\dotfill	& 0.04 & 0.08 & 0.11 & 0.00 & 0.06 & 0.04 & 0.09 & 0.03 & 0.10      \\
$\upsilon_{\rm turb}$ (km\,s$^{-1}$)\dotfill	&  10 &  15 &  12 & 15 & 12 & 15 & 15 & 10 & 15 \\
$Y_{\rm He}$\dotfill	&  0.08 &  0.08 &  0.10 & 0.08 & 0.08 & 0.10 & 0.08 & 0.10 & 0.08 \\
$\epsilon_{\rm C}$\dotfill	& $7.15\pm0.20$ & $<7.30$ & $<6.90$ & $7.20\pm0.20$ & \nodata & $6.95\pm0.20$ & $7.20\pm0.20$ & \nodata &
$7.00\pm0.20$ \\
$\epsilon_{\rm N}$\dotfill  	& $7.40\pm0.25$ & $7.10\pm0.20$ & $7.10\pm0.20$ & $<7.65$ & $<7.20$ & $7.00\pm0.20$ & $7.10\pm0.20$ & \nodata & $7.30\pm0.20$\\
$\epsilon_{\rm O}$\dotfill  	& $7.85\pm0.25$ & $7.85\pm0.17$ & $7.85\pm0.17$ & $7.90\pm0.20$ & $7.95\pm0.20$ & $8.00\pm0.20$ & $8.00\pm0.20$ &
\nodata & $7.80\pm0.20$\\
$\epsilon_{\rm Mg}$\dotfill 	& $6.80\pm0.20$ & $<6.90$ &  $<6.75$ & $<6.75$ & \nodata & $<6.60$ & \nodata & $6.90\pm0.20$ & $6.75\pm0.20$\\
$\epsilon_{\rm Si}$\dotfill 	& $6.65\pm0.20$ & $6.90\pm0.20$ & $6.65\pm0.20$ & $6.60\pm0.20$ & $6.65\pm0.20$ & $6.60\pm0.20$ & $6.60\pm0.20$ &
$6.60\pm0.20$ & $6.60\pm0.20$\\
$[{\rm O/H}]$\tablenotemark{c} (dex)\dotfill	&  $-0.85$ & $-0.85$ & $-0.85$ & $-0.80$ & $-0.75$ & $-0.70$ & $-0.70$ & \nodata & $-0.9$\\
\enddata
\tablecomments{Abundances are expressed as $\epsilon_{\rm X}=12+\log(X/H)$.}
\tablenotetext{a}{Error: $\pm1000$~K.}
\tablenotetext{b}{Error: $\pm0.10$.}
\tablenotetext{c}{Adopting $\epsilon_{\rm O, \odot}=8.66$ (\citealt{Asplund:2004}).}
\end{deluxetable*}

The modeling with {\sc fastwind} provides us with effective temperatures (\teff)
from the helium (late-O stars) and silicon (early-B stars) ionization
equilibria, while fits to the Balmer line profiles are used to determine surface
gravities ($\log\,g$). Among the additional parameters that are estimated by the
fitting procedure are the microturbulent velocity, $\upsilon_{\rm turb}$ (using
a combination of \hei\/ and the \siii/{\sc iii}/{\sc iv} lines) and the
abundances of various metals. The main abundance diagnostics used in this work
are \cii\llin6578--6583; \nii\llin3995, 5050, 5100; {\sc oii}\llin4072--4076,
4317--4319, 4414-4416; \mgii\lin4481; \siiii\llin4553--4575. One or more of
these features are not always available with sufficient S/N to provide strong
contraints on the chemical abundance. However, the large number of additional
faint metal lines and line blends that are present in the optical range,
especially originated by oxygen atoms, is used to determine the metal content
even when individual lines cannot be resolved in our spectra.  For the He
abundance we rely on \hei\llin4026, 4388, 4471, 4921, 5015, 5048 and 6678 (the
other lines  have uncertain broadening data). The reddening for each star has
been derived from a comparison of the observed {\em VI} and {\em JK} magnitudes
(the latter extracted from the near-IR photometry by \citealt{Pietrzynski:2006})
with the spectral energy distribution of the best-fitting {\sc fastwind} models,
and adopting the \citet{Cardelli:1989} extinction law with $R_V=3.1$. We obtain
rather low values for $E(B-V)$, with a mean value of $0.06 \pm 0.03$, 
similar to the result of $0.09 \pm 0.02$ obtained independently in our Cepheid study
(\citealt{Pietrzynski:2006}).

Our results are summarized in Table~\ref{modelparameters}, where the chemical
abundances of C, N, O, Mg and Si are expressed with the notation $\epsilon_{\rm
X}=12+\log(X/H)$. The distance-dependent parameters assume a distance to IC~1613
of 721~kpc (\citealt{Pietrzynski:2006}). Our estimated errors are 1000~K in
\teff, 0.1 dex in $\log g$, and vary between 0.17 dex and 0.25 dex in metal
abundances. Spectroscopic masses $M_{\rm spec}$ are lower limits, due to the
fact that at the FORS spectral resolution we cannot determine corrections to the
gravities that account for the stellar rotational velocities. The sensitivity of
the derived chemical abundances to the different stellar parameters for stars of
similar \teff\/ and from similar data as ours have been presented in detail by
\citet{Urbaneja:2005} and \citet{Evans:2007}, and will not be repeated here. In
Fig.~\ref{models1} and \ref{models2} we plot the adopted final models on top of
the observed spectra for the 9 B-type supergiants we have analyzed. Fig.~\ref{models4} presents 
a comparison between the spectrum of the B2~Iab star A7 in the 4300-4700~\AA\/ wavelength
range and {\sc fastwind} models calculated at the adopted metallicity (for oxygen: $\epsilon_O=7.85\pm0.25$, dotted line
in the center), and at 0.2 dex higher (top) and lower (bottom) metal abundances.
This example is a typical case: while overall the strength of the observed spectral features agree with the 
adopted model, some features (e.g.~\oii\llin4415-4417 for this star) are better 
reproduced with the 0.2 dex higher abundance, while others (e.g.~\oii\llin4317-4319 in this example) would suggest a slightly
lower abundance.

The mean oxygen abundance of the stars in Table~\ref{modelparameters} is
\oh~7.90~$\pm$~0.08. In Section~\ref{nebular} we obtain from two \hii\/ regions
an average nebular abundance \oh~7.73~$\pm$~0.03. In view of the errors, 
we regard the slight discrepancy as not significant. The median oxygen
abundance of the B supergiants of 7.9 compares to very similar values of 7.9
obtained in WLM (\citealt{Bresolin:2006}) and of 7.8 in NGC~3109
(\citealt{Evans:2007}).

Regarding the additional chemical elements, in several cases we could only
obtain upper limits to their abundances, except for silicon. Nitrogen abundances
could be measured for 6 stars. They are considerably higher than the \hii\/
region values of 12~+~log(N/H)~=~6.5--6.7 (\citealt{Lee:2003},
\citealt{Kingsburgh:1995}), by up to $\sim$0.7~dex. Nitrogen enrichments of
comparable size are found in samples of B supergiants in, for example, the SMC
(\citealt{Trundle:2004}). Excluding upper limits, the mean abundances (by number
of particles) of the remaining elements relative to oxygen, log(C/O)~=~$-0.8$,
log(Si/O)~=~$-1.20$ and log(Mg/O)~=~$-1.05$, are also comparable to the SMC B
supergiant values (\citealt{Dufton:2005}).

\section{Nebular abundances}\label{nebular}

Several of our stellar spectra show contamination by nebular lines, as
summarized in the last column of Table~\ref{catalog}. This offers us the
possibility of comparing \hii\/ region chemical abundances with the
metallicities derived for the supergiant stars. In order to determine accurate
nebular abundances, the knowledge of the electron temperature of the gas, \te,
is necessary, since the emissivities of the forbidden lines used for nebular
abundance work strongly depend on it. At low metallicity the auroral-to-nebular
line ratio \oiii~\lin4363/(\lin4959+\lin5007) can be used for such purpose.
However, the \hii\/ regions in IC~1613 are of low surface brightness, making it
difficult to detect the faint \lin4363 auroral line. Abundance measurements
based on this line have been carried out in the \hii\/ region S3
(\citealt{Sandage:1971}) by several authors, thanks to the high degree of
excitation of this nebula, ionized by the hot WO3 star described previously,
that makes the \oiii\/ lines very strong. In their early work
\citet{Davidson:1982} obtained an oxygen abundance \oh~7.87, while more modern
measurements by \citet{Kingsburgh:1995} and \citet{Lee:2003} are in the range
\oh~7.62--7.70. Work on additional \hii\/ regions has derived chemical
abundances from bright-line methods and comparisons of the line flux ratios to
photoionization model results, without a direct measurement of the electron
temperature. In particular, the chemical abundance for the supernova remnant S8
has been studied by \citet[\oh~7.60]{Dodorico:1983} and
\citet[\oh~7.83]{Peimbert:1988}. \citet{Lee:2003} determined \oh~7.90 for region
\#13 (\citealt{Hodge:1990}) from the R$_{23}$ method. A N/O ratio enhanced
relative to other low-luminosity dwarf irregular galaxies, in the range
N/O~=~0.07--0.12, is found in the papers cited.

\hoffset=0 mm
\begin{deluxetable}{cccc}
\tabletypesize{\scriptsize}
\tablecolumns{4}
\tablewidth{0pt}
\tablecaption{H\,II regions: reddening-corrected line fluxes\label{fluxes}}

\tablehead{
\colhead{\phantom{aaaaaa}$\lambda_0$ (\AA)\phantom{aaaaaa}}	 &
\colhead{Ion}	&
\colhead{S3}	&
\colhead{S17} }

\startdata
\\[-1mm]
3727\dotfill &  \oii   	&    93  $\pm$  5     		&    299 $\pm$    15	\\ 
3869\dotfill &  \neiii &       55 $\pm$ 3		&     10.9 $\pm$  0.9		\\ 
3889\dotfill &  \hi\/ + He\,I  &   18.6  $\pm$  0.9	&   20.4 $\pm$  1.2	\\ 
3969\dotfill &  \hi\/ + \neiii   & 34 $\pm$ 2 	&   22.7 $\pm$  1.4	\\ 
4026\dotfill &  \hei      &   1.8 $\pm$ 0.3		&   \nodata	\\ 
4072\dotfill &  \sii   &   1.2 $\pm$ 0.4		&   \nodata	\\ 
4101\dotfill &  \hi       &   27  $\pm$  1	&   26 $\pm$  1	\\ 
4340\dotfill &  \hi       &   50  $\pm$  2	&   49 $\pm$  2	\\ 
4363\dotfill &  \oiii  &   15.1 $\pm$ 0.6		&   3.9 $\pm$ 0.4		\\ 
4471\dotfill &  \hei   &   3.6  $\pm$  0.2		&    3.2 $\pm$  0.4	\\ 
4686\dotfill &  \heii      &   24.1 $\pm$ 0.9		&   \nodata	\\ 
4711\dotfill &  \ariv  &   3.5 $\pm$ 0.2		&   \nodata		\\ 
4740\dotfill &  \ariv  &   2.7 $\pm$ 0.2		&   \nodata		\\ 
4861\dotfill &  \hi       &   100  $\pm$    4	&    100 $\pm$    4	\\ 
4922\dotfill &  \hei      &   1.04 $\pm$ 0.32		&   \nodata	\\ 
4959\dotfill &  \oiii  &     183  $\pm$    7		&      73 $\pm$   3	\\ 
5007\dotfill &  \oiii  &     521  $\pm$   20		&      221 $\pm$  9	\\ 
5412\dotfill &  \heii      &   2.4 $\pm$ 0.2		&   \nodata	\\ 
5876\dotfill &  \hei      &   7.1  $\pm$  0.4		&    11.4 $\pm$  0.7	\\ 
\enddata
\tablecomments{Fluxes are normalized to H$\beta$~=~100.}
\end{deluxetable}

\begin{deluxetable}{lcc}
\tabletypesize{\scriptsize}
\tablecolumns{3}
\tablewidth{0pt}
\tablecaption{H\,II regions: derived parameters\label{parameters}}

\tablehead{
\colhead{\phantom{aaaaaaaaaaaaaaaaaaaaa}}	 &
\colhead{S3}	&
\colhead{S17}}

\startdata
\\[-1mm]
\toiii\dotfill			&	18300 $\pm$ 400 		&	14500 $\pm$ 700			\\
\toii\dotfill			&	14900 $\pm$ 50 			&	13600 $\pm$ 400			\\
O$^+$/H$^+$\dotfill		&	$(7.0 \pm 0.1) \times 10^{-6}$	&	$(3.4 \pm 0.3) \times 10^{-5}$	\\
O$^{++}$/H$^+$\dotfill		&	$(3.5 \pm 0.2) \times 10^{-5}$	&	$(2.6 \pm 0.3) \times 10^{-5}$	\\
He$^+$/H$^+$\dotfill		&	0.057 $\pm$ 0.004		&	0.092 $\pm$ 0.007		\\
He$^{++}$/H$^+$\dotfill		&	0.022 $\pm$ 0.001		&	\nodata				\\
ICF(O)\dotfill			&	1.24				&	1				\\
12\,+\,log(O/H)\dotfill		&	7.63 $\pm$ 0.02			&	7.78 $\pm$ 0.05			\\
12\,+\,log(O/H)$_c$\tablenotemark{a}\dotfill	&	7.72 $\pm$ 0.02			&	7.78 $\pm$ 0.05			\\
He/H\dotfill			&	0.079 $\pm$ 0.005		&	0.092 $\pm$ 0.007		\\
\enddata
\tablenotetext{a}{Oxygen abundance corrected for the ionization correction factor ICF(O).}
\end{deluxetable}

Two stars in our sample are associated with \hii\/ regions in which we have
detected \oiii\lin4363: B17 (\hii\/ region S3) and B2 (\hii\/ region S17). For
the spectral extraction of these two \hii\/ regions we have avoided the stars,
thus producing nebular spectra that are unaffected by strong stellar continua
(Fig.~\ref{spectrahii}). As already known since the early studies of the WO3
star DR1 and its associated nebula (\citealt{Dodorico:1982,Davidson:1982}), the
high-excitation spectrum of S3 contains the rare \heii\lin4686 line. In our
spectrum we have also detected the fainter \heii\/ \lin4542 and \lin5412 lines.

The emission line intensities, measured after flux-calibration of the spectra,
and normalized to I(H$\beta$)~=~100, are presented in Table~\ref{fluxes}. The
Balmer decrement, measured by the H$\gamma$/H$\beta$ and H$\delta$/H$\beta$ line
ratios, is consistent with a negligible amount of extinction for both \hii\/
regions. We have derived the electron temperature for the O$^{++}$-emitting
zone, \toiii, from the \oiii~\lin4363/(\lin4959+\lin5007) ratio and assuming the
low-density regime (N$_e$~=~100\,cm$^{-3}$), using the {\em nebular} package in
{\sc iraf}. Our value for S3, \te~=~18300~$\pm$~400~K, compares well with
\te~=~17910~K given by \citet{Lee:2003}, and is slightly higher than
\te~=~17100~$\pm$~500~K reported by \citet{Kingsburgh:1995}. The temperature of
the O$^{+}$-emitting zones, \toii, has been derived from \toiii\/ and the
relations published by \citet{Izotov:2006}. The oxygen ionic abundances,
O$^+$/H$^+$ (obtained from \oii\lin3727/H$\beta$) and O$^{++}$/H$^+$ (from
\oiii\lin5007/H$\beta$), together with additional parameters, are summarized in
Table~\ref{parameters}.

For the \hii\/ region S17 the total oxygen abundance is given as the sum
O$^+$/H$^+$\,+\,O$^{++}$/H$^+$, and we obtain \oh~7.78~$\pm$~0.05. For the high
excitation region S3 we need to account for the contribution of O$^{3+}$, which
is not observed in the optical range. To estimate an ionization correction
factor (ICF) we have followed \citet{Kingsburgh:1994}:
ICF(O)~=~(1~+~He$^{++}$/He$^+$)$^{2/3}$~=~1.24. With this correction, the oxygen
abundance of S3 becomes \oh~7.72~$\pm$~0.02. This value agrees with the result
by \citet[\oh~7.70]{Kingsburgh:1995}. \citet{Lee:2003} obtained
\oh~7.62~$\pm$~0.05, but they did not apply any correction to account for the
presence of O$^{3+}$. Our uncorrected value is \oh~7.63~$\pm$~0.02.

To summarize, besides the well-studied \hii\/ region S3, ionized by the WO3 star
DR1, with this study we have found a second \hii\/ region in IC~1613 in which we
have a direct measurement of the electron temperature. The oxygen abundances of
two \hii\/ regions  agree quite well, with an average \oh~7.73~$\pm$~0.04. We
also note the good agreement found between nebular and stellar helium
abundances: the mean from the B supergiants is He/H~=~0.087~$\pm$~0.010, while
the weighted mean from the \hii\/ regions is He/H~=~0.083~$\pm$~0.009,

\section{Summary}

In this paper we have presented multi-object spectroscopy of young, massive
stars that we have obtained in the Local Group galaxy IC~1613. We have provided
the spectral classification and a detailed spectral catalog for 54 OBA stars in
this galaxy. The majority of the photometrically selected sample is composed of
B- and A-type supergiants. The remaining stars include early O-type dwarfs and
the only Wolf-Rayet star known in this galaxy. Among the early B stars we have
uncovered 6 Be stars, which constitute the largest spectroscopically confirmed
sample of this class of objects beyond the Magellanic Clouds. The radial
velocities of all these stars is consistent with the systemic velocity. Only
three stars have clearly discrepant velocities, and are identified as foreground
objects belonging to the Milky Way. Of these, 2 G-type stars clearly violate the
criterion used for the selection of blue supergiants. The third one, an A star
in the Galactic halo, has very weak metal lines, and would have been flagged as
peculiar even in the absence of a discrepant radial velocity information.

Using model atmospheres calculated with {\sc fastwind} we have measured chemical
abundances for 9 early-B supergiants, and have found a mean oxygen abundance of
12~+~log(O/H)~=~7.90~$\pm$~0.08. This value is comparable with the result we
obtain for two \hii\/ regions in which we detect the temperature-sensitive
\oiii\lin4363 auroral line, \oh~7.73~$\pm$~0.04. The agreement extends to the
helium abundance. The abundance patterns we find for the remaining chemical
elements is similar to those measured in B supergiants in the slightly more
metal-rich SMC. The stellar oxygen abundance is very close to the values we have
found from similar data in the other dwarf galaxies whose blue supergiants have
been studied so far as part of the Araucaria project, WLM and NGC 3109.

\acknowledgments FB would like to thank the ESO-Paranal staff for their high
standard of support throughout this project. We thank A.~Herrero for providing
us with the {\em UBV} photometry. GP and WG gratefully acknowledge financial
support for this work from the Chilean Center for Astrophysics, under grant
FONDAP 15010003.


\begin{thebibliography}{68}
\expandafter\ifx\csname natexlab\endcsname\relax\def\natexlab#1{#1}\fi

\bibitem[{{Armandroff} \& {Massey}(1991)}]{Armandroff:1991}
{Armandroff}, T.~E. \& {Massey}, P. 1991, \aj, 102, 927

\bibitem[{{Arnold} \& {Gilmore}(1992)}]{Arnold:1992}
{Arnold}, R. \& {Gilmore}, G. 1992, \mnras, 257, 225

\bibitem[{{Asplund} {et~al.}(2004){Asplund}, {Grevesse}, {Sauval}, {Allende
  Prieto}, \& {Kiselman}}]{Asplund:2004}
{Asplund}, M., {Grevesse}, N., {Sauval}, A.~J., {Allende Prieto}, C., \&
  {Kiselman}, D. 2004, \aap, 417, 751

\bibitem[{{Azzopardi}(1987)}]{Azzopardi:1987}
{Azzopardi}, M. 1987, \aaps, 69, 421

\bibitem[{{Azzopardi} {et~al.}(1988){Azzopardi}, {Lequeux}, \&
  {Maeder}}]{Azzopardi:1988}
{Azzopardi}, M., {Lequeux}, J., \& {Maeder}, A. 1988, \aap, 189, 34

\bibitem[{{Borissova} {et~al.}(2004){Borissova}, {Kurtev}, {Georgiev}, \&
  {Rosado}}]{Borissova:2004}
{Borissova}, J., {Kurtev}, R., {Georgiev}, L., \& {Rosado}, M. 2004, \aap, 413,
  889

\bibitem[{{Bresolin} {et~al.}(2002){Bresolin}, {Gieren}, {Kudritzki},
  {Pietrzy{\'n}ski}, \& {Przybilla}}]{Bresolin:2002}
{Bresolin}, F., {Gieren}, W., {Kudritzki}, R.-P., {Pietrzy{\'n}ski}, G., \&
  {Przybilla}, N. 2002, \apj, 567, 277

\bibitem[{{Bresolin} {et~al.}(2006){Bresolin}, {Pietrzy{\'n}ski}, {Urbaneja},
  {Gieren}, {Kudritzki}, \& {Venn}}]{Bresolin:2006}
{Bresolin}, F., {Pietrzy{\'n}ski}, G., {Urbaneja}, M.~A., {Gieren}, W.,
  {Kudritzki}, R.-P., \& {Venn}, K.~A. 2006, \apj, 648, 1007

\bibitem[{{Cardelli} {et~al.}(1989){Cardelli}, {Clayton}, \&
  {Mathis}}]{Cardelli:1989}
{Cardelli}, J.~A., {Clayton}, G.~C., \& {Mathis}, J.~S. 1989, \apj, 345, 245

\bibitem[{{Cole} {et~al.}(1999){Cole}, {Tolstoy}, {Gallagher}, {Hoessel},
  {Mould}, {Holtzman}, {Saha}, {Ballester}, {Burrows}, {Clarke}, {Crisp},
  {Griffiths}, {Grillmair}, {Hester}, {Krist}, {Meadows}, {Scowen},
  {Stapelfeldt}, {Trauger}, {Watson}, \& {Westphal}}]{Cole:1999}
{Cole}, A.~A., {Tolstoy}, E., {Gallagher}, III, J.~S., {Hoessel}, J.~G.,
  {Mould}, J.~R., {Holtzman}, J.~A., {Saha}, A., {Ballester}, G.~E., {Burrows},
  C.~J., {Clarke}, J.~T., {Crisp}, D., {Griffiths}, R.~E., {Grillmair}, C.~J.,
  {Hester}, J.~J., {Krist}, J.~E., {Meadows}, V., {Scowen}, P.~A.,
  {Stapelfeldt}, K.~R., {Trauger}, J.~T., {Watson}, A.~M., \& {Westphal}, J.~R.
  1999, \aj, 118, 1657

\bibitem[{{Conti}(1988)}]{Conti:1988}
{Conti}, P.~S. 1988, {in O stars and Wolf-Rayet stars, ed. P.S. Conti \& A.B.
  Underhill, NASA SP-497}

\bibitem[{{Conti} \& {Leep}(1974)}]{Conti:1974}
{Conti}, P.~S. \& {Leep}, E.~M. 1974, \apj, 193, 113

\bibitem[{{Davidson} \& {Kinman}(1982)}]{Davidson:1982}
{Davidson}, K. \& {Kinman}, T.~D. 1982, \pasp, 94, 634

\bibitem[{{D'Odorico} \& {Dopita}(1983)}]{Dodorico:1983}
{D'Odorico}, S. \& {Dopita}, M. 1983, in IAU Symp. 101: Supernova Remnants and
  their X-ray Emission, ed. J.~{Danziger} \& P.~{Gorenstein}, 517--524

\bibitem[{{D'Odorico} \& {Rosa}(1982)}]{Dodorico:1982}
{D'Odorico}, S. \& {Rosa}, M. 1982, \aap, 105, 410

\bibitem[{{Dufton} {et~al.}(2005){Dufton}, {Ryans}, {Trundle}, {Lennon},
  {Hubeny}, {Lanz}, \& {Allende Prieto}}]{Dufton:2005}
{Dufton}, P.~L., {Ryans}, R.~S.~I., {Trundle}, C., {Lennon}, D.~J., {Hubeny},
  I., {Lanz}, T., \& {Allende Prieto}, C. 2005, \aap, 434, 1125

\bibitem[{{Evans} {et~al.}(2007){Evans}, {Bresolin}, {Urbaneja},
  {Pietrzy{\'n}ski}, {Gieren}, \& {Kudritzki}}]{Evans:2007}
{Evans}, C.~J., {Bresolin}, F., {Urbaneja}, M.~A., {Pietrzy{\'n}ski}, G.,
  {Gieren}, W., \& {Kudritzki}, R.-P. 2007, \apj, 659, 1198

\bibitem[{{Evans} \& {Howarth}(2003)}]{Evans:2003}
{Evans}, C.~J. \& {Howarth}, I.~D. 2003, \mnras, 345, 1223

\bibitem[{{Evans} {et~al.}(2004){Evans}, {Howarth}, {Irwin}, {Burnley}, \&
  {Harries}}]{Evans:2004}
{Evans}, C.~J., {Howarth}, I.~D., {Irwin}, M.~J., {Burnley}, A.~W., \&
  {Harries}, T.~J. 2004, \mnras, 353, 601

\bibitem[{{Freedman}(1988)}]{Freedman:1988}
{Freedman}, W.~L. 1988, \aj, 96, 1248

\bibitem[{{Garnett} {et~al.}(1991){Garnett}, {Kennicutt}, {Chu}, \&
  {Skillman}}]{Garnett:1991}
{Garnett}, D.~R., {Kennicutt}, Jr., R.~C., {Chu}, Y.-H., \& {Skillman}, E.~D.
  1991, \apj, 373, 458

\bibitem[{{Georgiev} {et~al.}(1999){Georgiev}, {Borissova}, {Rosado}, {Kurtev},
  {Ivanov}, \& {Koenigsberger}}]{Georgiev:1999}
{Georgiev}, L., {Borissova}, J., {Rosado}, M., {Kurtev}, R., {Ivanov}, G., \&
  {Koenigsberger}, G. 1999, \aaps, 134, 21

\bibitem[{{Gieren} {et~al.}(2005){Gieren}, {Pietrzynski}, {Bresolin},
  {Kudritzki}, {Minniti}, {Urbaneja}, {Soszynski}, {Storm}, {Fouque}, {Bono},
  {Walker}, \& {Garcia}}]{Gieren:2005}
{Gieren}, W., {Pietrzynski}, G., {Bresolin}, F., {Kudritzki}, R.-P., {Minniti},
  D., {Urbaneja}, M., {Soszynski}, I., {Storm}, J., {Fouque}, P., {Bono}, G.,
  {Walker}, A., \& {Garcia}, J. 2005, The Messenger, 121, 23

\bibitem[{{Hodge} {et~al.}(1990){Hodge}, {Lee}, \& {Gurwell}}]{Hodge:1990}
{Hodge}, P., {Lee}, M.~G., \& {Gurwell}, M. 1990, \pasp, 102, 1245

\bibitem[{{Hodge}(1978)}]{Hodge:1978}
{Hodge}, P.~W. 1978, \apjs, 37, 145

\bibitem[{{Hodge} {et~al.}(1991){Hodge}, {Smith}, {Eskridge}, {MacGillivray},
  \& {Beard}}]{Hodge:1991}
{Hodge}, P.~W., {Smith}, T.~R., {Eskridge}, P.~B., {MacGillivray}, H.~T., \&
  {Beard}, S.~M. 1991, \apj, 369, 372

\bibitem[{{Hoffman} {et~al.}(1996){Hoffman}, {Salpeter}, {Farhat}, {Roos},
  {Williams}, \& {Helou}}]{Hoffman:1996}
{Hoffman}, G.~L., {Salpeter}, E.~E., {Farhat}, B., {Roos}, T., {Williams}, H.,
  \& {Helou}, G. 1996, \apjs, 105, 269

\bibitem[{{Humphreys}(1980)}]{Humphreys:1980}
{Humphreys}, R.~M. 1980, \apj, 238, 65

\bibitem[{{Izotov} {et~al.}(2006){Izotov}, {Stasi{\'n}ska}, {Meynet}, {Guseva},
  \& {Thuan}}]{Izotov:2006}
{Izotov}, Y.~I., {Stasi{\'n}ska}, G., {Meynet}, G., {Guseva}, N.~G., \&
  {Thuan}, T.~X. 2006, \aap, 448, 955

\bibitem[{{Jaschek} {et~al.}(1980){Jaschek}, {Jaschek}, {Hubert-Delplace}, \&
  {Hubert}}]{Jaschek:1980}
{Jaschek}, M., {Jaschek}, C., {Hubert-Delplace}, A.-M., \& {Hubert}, H. 1980,
  \aaps, 42, 103

\bibitem[{{Kingsburgh} \& {Barlow}(1994)}]{Kingsburgh:1994}
{Kingsburgh}, R.~L. \& {Barlow}, M.~J. 1994, \mnras, 271, 257

\bibitem[{{Kingsburgh} \& {Barlow}(1995)}]{Kingsburgh:1995}
---. 1995, \aap, 295, 171

\bibitem[{{Lake} \& {Skillman}(1989)}]{Lake:1989}
{Lake}, G. \& {Skillman}, E.~D. 1989, \aj, 98, 1274

\bibitem[{{Lee} {et~al.}(2003){Lee}, {Grebel}, \& {Hodge}}]{Lee:2003}
{Lee}, H., {Grebel}, E.~K., \& {Hodge}, P.~W. 2003, \aap, 401, 141

\bibitem[{{Lennon}(1997)}]{Lennon:1997}
{Lennon}, D.~J. 1997, \aap, 317, 871

\bibitem[{{Lequeux} {et~al.}(1987){Lequeux}, {Meyssonnier}, \&
  {Azzopardi}}]{Lequeux:1987}
{Lequeux}, J., {Meyssonnier}, N., \& {Azzopardi}, M. 1987, \aaps, 67, 169

\bibitem[{{Lozinskaya} {et~al.}(2002){Lozinskaya}, {Arkhipova}, {Moiseev}, \&
  {Afanas'Ev}}]{Lozinskaya:2002}
{Lozinskaya}, T.~A., {Arkhipova}, V.~P., {Moiseev}, A.~V., \& {Afanas'Ev},
  V.~L. 2002, Astronomy Reports, 46, 16

\bibitem[{{Lozinskaya} {et~al.}(2003){Lozinskaya}, {Moiseev}, \&
  {Podorvanyuk}}]{Lozinskaya:2003}
{Lozinskaya}, T.~A., {Moiseev}, A.~V., \& {Podorvanyuk}, N.~Y. 2003, Astronomy
  Letters, 29, 77

\bibitem[{{Lu} {et~al.}(1993){Lu}, {Hoffman}, {Groff}, {Roos}, \&
  {Lamphier}}]{Lu:1993}
{Lu}, N.~Y., {Hoffman}, G.~L., {Groff}, T., {Roos}, T., \& {Lamphier}, C. 1993,
  \apjs, 88, 383

\bibitem[{{Maeder} {et~al.}(1999){Maeder}, {Grebel}, \&
  {Mermilliod}}]{Maeder:1999}
{Maeder}, A., {Grebel}, E.~K., \& {Mermilliod}, J.-C. 1999, \aap, 346, 459

\bibitem[{{Martayan} {et~al.}(2007){Martayan}, {Fr{\'e}mat}, {Hubert},
  {Floquet}, {Zorec}, \& {Neiner}}]{Martayan:2007}
{Martayan}, C., {Fr{\'e}mat}, Y., {Hubert}, A.-M., {Floquet}, M., {Zorec}, J.,
  \& {Neiner}, C. 2007, \aap, 462, 683

\bibitem[{{Martins} {et~al.}(2005){Martins}, {Schaerer}, \&
  {Hillier}}]{Martins:2005}
{Martins}, F., {Schaerer}, D., \& {Hillier}, D.~J. 2005, \aap, 436, 1049

\bibitem[{{McConnachie} \& {Irwin}(2006)}]{McConnachie:2006}
{McConnachie}, A.~W. \& {Irwin}, M.~J. 2006, \mnras, 365, 902

\bibitem[{{Meaburn} {et~al.}(1988){Meaburn}, {Clayton}, \&
  {Whitehead}}]{Meaburn:1988}
{Meaburn}, J., {Clayton}, C.~A., \& {Whitehead}, M.~J. 1988, \mnras, 235, 479

\bibitem[{{Peimbert} {et~al.}(1988){Peimbert}, {Bohigas}, \&
  {Torres-Peimbert}}]{Peimbert:1988}
{Peimbert}, M., {Bohigas}, J., \& {Torres-Peimbert}, S. 1988, Revista Mexicana
  de Astronomia y Astrofisica, 16, 45

\bibitem[{{Pietrzy{\'n}ski} {et~al.}(2006){Pietrzy{\'n}ski}, {Gieren},
  {Soszy{\'n}ski}, {Bresolin}, {Kudritzki}, {Dall'Ora}, {Storm}, \&
  {Bono}}]{Pietrzynski:2006}
{Pietrzy{\'n}ski}, G., {Gieren}, W., {Soszy{\'n}ski}, I., {Bresolin}, F.,
  {Kudritzki}, R.-P., {Dall'Ora}, M., {Storm}, J., \& {Bono}, G. 2006, \apj,
  642, 216

\bibitem[{{Puls} {et~al.}(2005){Puls}, {Urbaneja}, {Venero}, {Repolust},
  {Springmann}, {Jokuthy}, \& {Mokiem}}]{Puls:2005}
{Puls}, J., {Urbaneja}, M.~A., {Venero}, R., {Repolust}, T., {Springmann}, U.,
  {Jokuthy}, A., \& {Mokiem}, M.~R. 2005, \aap, 435, 669

\bibitem[{{Rizzi} {et~al.}(2007){Rizzi}, {Tully}, {Makarov}, {Makarova},
  {Dolphin}, {Sakai}, \& {Shaya}}]{Rizzi:2007}
{Rizzi}, L., {Tully}, R.~B., {Makarov}, D., {Makarova}, L., {Dolphin}, A.~E.,
  {Sakai}, S., \& {Shaya}, E.~J. 2007, ArXiv Astrophysics e-prints

\bibitem[{{Sandage}(1971)}]{Sandage:1971}
{Sandage}, A. 1971, \apj, 166, 13

\bibitem[{{Sandage} \& {Katem}(1976)}]{Sandage:1976}
{Sandage}, A. \& {Katem}, B. 1976, \aj, 81, 743

\bibitem[{{Santolaya-Rey} {et~al.}(1997){Santolaya-Rey}, {Puls}, \&
  {Herrero}}]{Santolaya-Rey:1997}
{Santolaya-Rey}, A.~E., {Puls}, J., \& {Herrero}, A. 1997, \aap, 323, 488

\bibitem[{{Schlegel} {et~al.}(1998){Schlegel}, {Finkbeiner}, \&
  {Davis}}]{Schlegel:1998}
{Schlegel}, D.~J., {Finkbeiner}, D.~P., \& {Davis}, M. 1998, \apj, 500, 525

\bibitem[{{Schmidt-Kaler}(1982)}]{Schmidt-Kaler:1982}
{Schmidt-Kaler}, T. 1982, {Landolt-B{\"o}rnstein: Numerical Data and Functional
  Relationships in Science and Technology}, Vol.~VI (K. Schaifers and H.H.
  Voigt (Springer-Verlag, Berlin))

\bibitem[{{Silich} {et~al.}(2006){Silich}, {Lozinskaya}, {Moiseev},
  {Podorvanuk}, {Rosado}, {Borissova}, \& {Valdez-Gutierrez}}]{Silich:2006}
{Silich}, S., {Lozinskaya}, T., {Moiseev}, A., {Podorvanuk}, N., {Rosado}, M.,
  {Borissova}, J., \& {Valdez-Gutierrez}, M. 2006, \aap, 448, 123

\bibitem[{{Skillman} {et~al.}(2003){Skillman}, {Tolstoy}, {Cole}, {Dolphin},
  {Saha}, {Gallagher}, {Dohm-Palmer}, \& {Mateo}}]{Skillman:2003}
{Skillman}, E.~D., {Tolstoy}, E., {Cole}, A.~A., {Dolphin}, A.~E., {Saha}, A.,
  {Gallagher}, J.~S., {Dohm-Palmer}, R.~C., \& {Mateo}, M. 2003, \apj, 596, 253

\bibitem[{{Trundle} {et~al.}(2004){Trundle}, {Lennon}, {Puls}, \&
  {Dufton}}]{Trundle:2004}
{Trundle}, C., {Lennon}, D.~J., {Puls}, J., \& {Dufton}, P.~L. 2004, \aap, 417,
  217

\bibitem[{{Udalski} {et~al.}(2001){Udalski}, {Wyrzykowski}, {Pietrzynski},
  {Szewczyk}, {Szymanski}, {Kubiak}, {Soszynski}, \& {Zebrun}}]{Udalski:2001}
{Udalski}, A., {Wyrzykowski}, L., {Pietrzynski}, G., {Szewczyk}, O.,
  {Szymanski}, M., {Kubiak}, M., {Soszynski}, I., \& {Zebrun}, K. 2001, Acta
  Astronomica, 51, 221

\bibitem[{{Urbaneja} {et~al.}(2003){Urbaneja}, {Herrero}, {Bresolin},
  {Kudritzki}, {Gieren}, \& {Puls}}]{Urbaneja:2003}
{Urbaneja}, M.~A., {Herrero}, A., {Bresolin}, F., {Kudritzki}, R.-P., {Gieren},
  W., \& {Puls}, J. 2003, \apjl, 584, L73

\bibitem[{{Urbaneja} {et~al.}(2005){Urbaneja}, {Herrero}, {Bresolin},
  {Kudritzki}, {Gieren}, {Puls}, {Przybilla}, {Najarro}, \&
  {Pietrzy{\'n}ski}}]{Urbaneja:2005}
{Urbaneja}, M.~A., {Herrero}, A., {Bresolin}, F., {Kudritzki}, R.-P., {Gieren},
  W., {Puls}, J., {Przybilla}, N., {Najarro}, F., \& {Pietrzy{\'n}ski}, G.
  2005, \apj, 622, 862

\bibitem[{{Valdez-Guti{\'e}rrez} {et~al.}(2001){Valdez-Guti{\'e}rrez},
  {Rosado}, {Georgiev}, {Borissova}, \& {Kurtev}}]{Valdez-Gutierrez:2001}
{Valdez-Guti{\'e}rrez}, M., {Rosado}, M., {Georgiev}, L., {Borissova}, J., \&
  {Kurtev}, R. 2001, \aap, 366, 35

\bibitem[{{van Dokkum}(2001)}]{van-Dokkum:2001}
{van Dokkum}, P.~G. 2001, \pasp, 113, 1420

\bibitem[{{Walborn}(1971)}]{Walborn:1971}
{Walborn}, N.~R. 1971, \apjs, 23, 257

\bibitem[{{Walborn} \& {Fitzpatrick}(1990)}]{Walborn:1990}
{Walborn}, N.~R. \& {Fitzpatrick}, E.~L. 1990, \pasp, 102, 379

\bibitem[{{Walborn} {et~al.}(2000){Walborn}, {Lennon}, {Heap}, {Lindler},
  {Smith}, {Evans}, \& {Parker}}]{Walborn:2000}
{Walborn}, N.~R., {Lennon}, D.~J., {Heap}, S.~R., {Lindler}, D.~J., {Smith},
  L.~J., {Evans}, C.~J., \& {Parker}, J.~W. 2000, \pasp, 112, 1243

\bibitem[{{Weaver} {et~al.}(1977){Weaver}, {McCray}, {Castor}, {Shapiro}, \&
  {Moore}}]{Weaver:1977}
{Weaver}, R., {McCray}, R., {Castor}, J., {Shapiro}, P., \& {Moore}, R. 1977,
  \apj, 218, 377

\bibitem[{{Wilhelm} {et~al.}(1999{\natexlab{a}}){Wilhelm}, {Beers}, \&
  {Gray}}]{Wilhelm:1999b}
{Wilhelm}, R., {Beers}, T.~C., \& {Gray}, R.~O. 1999{\natexlab{a}}, \aj, 117,
  2308

\bibitem[{{Wilhelm} {et~al.}(1999{\natexlab{b}}){Wilhelm}, {Beers},
  {Sommer-Larsen}, {Pier}, {Layden}, {Flynn}, {Rossi}, \&
  {Christensen}}]{Wilhelm:1999a}
{Wilhelm}, R., {Beers}, T.~C., {Sommer-Larsen}, J., {Pier}, J.~R., {Layden},
  A.~C., {Flynn}, C., {Rossi}, S., \& {Christensen}, P.~R. 1999{\natexlab{b}},
  \aj, 117, 2329

\bibitem[{{Wisniewski} \& {Bjorkman}(2006)}]{Wisniewski:2006}
{Wisniewski}, J.~P. \& {Bjorkman}, K.~S. 2006, \apj, 652, 458

\end{thebibliography}


\clearpage

\clearpage

\end{document}